%% file: GreenTeamPaper_ArXiv.tex
\documentclass[final,5p,times]{elsarticle}

\usepackage[english]{babel}     
\usepackage[T1]{fontenc}        
\usepackage[utf8]{inputenc}     

\usepackage{eurosym}            

\usepackage{amsmath,amsfonts,amssymb,bm}
\usepackage[binary-units = true]{siunitx}
\DeclareSIUnit{\EUR}{\text{\euro}}
\DeclareSIUnit[number-unit-product = {}]{\degree}{\SIUnitSymbolDegree}

\usepackage{amsthm}
\makeatletter
    \def\els@aparagraph[#1]#2{\elsparagraph[#1]{#2\@addpunct{.}}}
    \def\els@bparagraph#1{\elsparagraph*{#1\@addpunct{.}}}
\makeatother

\usepackage[activate={true,nocompatibility},final,tracking=true,kerning=true,spacing=true,factor=1100,stretch=10,shrink=10]{microtype}
\microtypecontext{spacing=nonfrench}

\usepackage{graphicx}   
\usepackage{caption}

\usepackage{epstopdf}   

\usepackage{booktabs}   
\usepackage{multirow}   
\usepackage{threeparttable} 
\usepackage{arydshln}   

\usepackage[switch]{lineno}
\modulolinenumbers[5]

\usepackage[pdftex]{hyperref}

\usepackage{etoolbox}
\makeatletter
\patchcmd\@combinedblfloats{\box\@outputbox}{\unvbox\@outputbox}{}{\errmessage{\noexpand patch failed}}
\makeatother

\hypersetup{colorlinks          = true,         
            allcolors           = black,        
            bookmarksnumbered   = true,         
            bookmarksopen       = true,         
            breaklinks          = true,         
            pdfdisplaydoctitle  = true,         
            pdfpagelayout       = OneColumn,    
            pdfstartview        = FitH,         
            pdfpagemode         = UseNone,      
            pdftitle            = {Magrathea: Dust growth experiment in micro-gravity conditions},
            pdfauthor           = {Andre Guerra, Adrian Banos García, Adrian Castanon Esteban, Fabio Fabozzi, Marta Goli, Jonas Greif, Anton B. Ivanov, Lisa Jonsson, Kieran Leschinski, Victoria Lofstad, Marine Martin-Lagarde, John McClean, Mattia Reganaz, Julia Seibezeder, Esmee Stoop, Gwenael Van Looveren, Jophiel Wiis},
            pdfsubject          = {Article on the work developed by team green during the 2017 Alpbach Summer School.}
} 

\journal{ }

\biboptions{sort&compress}

\newcommand{\Figure}[1]{Figure~\ref{#1}}    
\newcommand{\Table}[1]{Table~\ref{#1}}  
\newcommand{\RSection}[1]{Section~\ref{#1}} 
\newcommand{\REF}[1]{\cite{#1}} 

\newcommand{\dustexperimentsSIZE}{1\columnwidth} 
\newcommand{\collisiontypesSIZE}{1\columnwidth} 
\newcommand{\payloadoverviewSIZE}{0.8\textwidth} 
\newcommand{\MagratheaSIZE}{0.7\columnwidth} 

\usepackage{xcolor}         

\usepackage[color=blue!20!white,linecolor=red,figcolor=white,textwidth=3cm,colorinlistoftodos,%
]{todonotes}                                                

\begin{document}


\begin{frontmatter}
    \title{Magrathea: Dust growth experiment in micro-gravity conditions}

    \author[ANDRE]{Andr\'{e} G. C. Guerra\corref{cor1}}
    \ead{andre.gc.guerra@gmail.com}

    \author[ADRIANGARCIA]{Adri\'{a}n Banos Garc\'{i}a}
    \ead{adrian.bgarci@gmail.com}
    
    \author[ADRIANESTEBAN]{Adri\'{a}n Castan\'{o}n Esteban}
    \ead{Adrian.Castanon-Esteban@student.uibk.ac.at}
    
    \author[FABIO]{Fabio Fabozzi}
    \ead{fabio.fabozzi1@gmail.com}
    
    \author[MARTA]{Marta Goli}
    \ead{martadgoli@gmail.com}
    
    \author[JONAS]{Jonas Greif}
    \ead{jonas.greif@uni-jena.de}
    
    \author[ANTON1,ANTON2]{Anton B.Ivanov}
    \ead{a.ivanov2@epfl.ch}
    
    \author[LISA]{Lisa Jonsson}
    \ead{lisa.cornelia.jonsson@gmail.com}
    
    \author[KIERAN]{Kieran Leschinski}
    \ead{kdleschinski@gmail.com}
    
    \author[VICTORIA]{Victoria Lofstad}
    \ead{victoria.lofstad@gmail.com}
    
    \author[MARINE]{Marine Martin-Lagarde}
    \ead{m.martin.lagarde@gmail.com}
    
    \author[JOHN]{John McClean}
    \ead{j.mcclean15@imperial.ac.uk}
    
    \author[MATTIA]{Mattia Reganaz}
    \ead{mattiareganaz@gmail.com}
    
    \author[JULIA]{Julia Seibezeder}
    \ead{julia.seibezeder@edu.uni-graz.at}
    
    \author[ESMEE]{Esmee Stoop}
    \ead{stoopesmee@gmail.com}

    \author[GWENAEL]{Gwena{\"e}l Van Looveren}
    \ead{gwenael.vl@gmail.com}

    \author[JOPHIEL]{Jophiel Wiis}
    \ead{jophiel@gmail.com}

    \cortext[cor1]{Corresponding author}

    \address[ANDRE]{Departamento de F\'{i}sica e Astronomia, Centro de F\'{i}sica do Porto, Faculdade de Ci\^{e}ncias, Universidade do Porto, Portugal}

    \address[ADRIANGARCIA]{Universidad Politécnica de Madrid, Espa\~na}
    
    \address[ADRIANESTEBAN]{Universität Innsbruck, Austria}
    
    \address[FABIO]{University of Rome La Sapienza, Italy}
    
    \address[MARTA]{Warsaw University of Technology, Poland}
    
    \address[JONAS]{University Jena, Astrophysical Institute and University-Observatory, Germany}
    
    \address[ANTON1]{EPFL Space Center, Ecole Polytechnique Federale de Lausanne, Lausanne, Switzerland}
    \address[ANTON2]{Space Center, Skolkovo Institute of Science and Technology, Moscow, Russia}
    
    \address[LISA]{Lule\r{a} University of Technology, Sweden}
    
    \address[KIERAN]{University of Vienna, Department of Astrophysics, Austria}
    
    \address[VICTORIA]{University of Oslo, Norway}
    
    \address[MARINE]{CEA Paris-Saclay, France}
    
    \address[JOHN]{Imperial College London, United Kingdom}
    
    \address[MATTIA]{Chair of Turbomachinery and Flight Propulsion, Technical University of Munich, Germany}
    
    \address[JULIA]{IGAM, Karl-Franzens-Universität Graz, Austria}
    
    \address[ESMEE]{Leiden Observatory, Faculty of Science of Leiden University, The Netherlands}

    \address[GWENAEL]{KU Leuven, Belgium}

    \address[JOPHIEL]{University of Copenhagen, Denmark}
    
    \begin{abstract}
        One of the least understood processes in astrophysics is the formation of planetesimals from molecules and dust within protoplanetary disks.
        In fact, current methods have strong limitations when it comes to model the full dynamics in this phase of planet formation, where small dust aggregates collide and grow into bigger clusters.
        That is why microgravity experiments of the phenomena involved are important to reveal the underlying physics.
        Because previous experiments had some limitations, in particular short durations and constrained dimensions, a new mission to study the very first stages of planet formation is proposed here.
        This mission, called Magrathea, is focused on creating the best conditions for developing these experiments, using a satellite with a \SI{6}{m^3} test chamber.
        During the mission 28 experiments are performed using different dust compositions, sizes and shapes, to better understand under which conditions dust grains stick and aggregate.
        Each experiment should last up to one month, with relative collision velocities of up to \SI{5}{\milli \metre \per \second}, and initial dust sizes between \SI{1}{\micro \metre} and \SI{1}{\milli \metre}.
        At least $10^6$ collisions per experiment should be recorded, to provide statistically significant results.
        Based on the scientific objectives and requirements, a preliminary analysis of the payload instrumentation is performed.
        From that a conceptual mission and spacecraft design is developed, together with a first approach to mission programmatic and risk analysis.
        The solution reached is a \SI{1000}{kg} spacecraft, set on a \SI{800}{km} Sun-synchronous orbit, with a total mission cost of around \SI{438}{MEuros}.
    \end{abstract}


\end{frontmatter}



\input{1_ScienceBackground}
\input{2_ScienceObjectives}
\input{4_MissionDesign}

\input{3_Payload}

\input{5_Spacecraft}

\input{7_Programmatic}

\input{8_Conclusion}


\section*{Acknowledgements}

The authors acknowledge funding from ESA and the Austrian Research Promotion Agency during the Alpbach Summer School 2017.
The attendance of Marta Goli to the Summer School was partially supported by Europlanet 2020 RI.
The attendance of Andr\'{e} Guerra to the Summer School was partially supported by Europlanet 2020 RI and Centro de F\'{i}sica do Porto.
The work of Andr\'{e} Guerra is supported by the Funda{\c c}\~{a}o para a Ci\^{e}ncia e a Tecnologia (Portuguese Agency for Research) fellowship PD/BD/113536/2015.
The attendance of Anton Ivanov was supported by the Swiss Committee on Space Research.
The authors would like to thank the organisers of the Alpbach Summer School for all the support before, during and after the school, in particular to Michaela Gitsch.
The authors would also like to thank all Summer School tutors and lecturers for their many inputs into the subject.
The authors acknowledge the help of Carsten Scharlemann and Orfeu Bertolami for reviewing this paper, and of Inga Kamp for her publishing suggestions.

\bibliography{GreenTeamPaper_BIB}

%

\end{document}

%% file: 1_ScienceBackground.tex
%
%
%

\section{Scientific background -- Grain growth process and its study}
\label{sec:Science_Background}

Recent observations of exoplanetary systems have revealed a wide variety of planets in the Universe~\REF{Borucki2010,Broeg2013,Knutson2007}.
While the processes involved in planet formation from planetesimals are relatively well researched, the ones that allow for dust particles to grow into planetesimals are poorly understood.
Moreover, the theories and simulations of grain growth from micron- to kilometre-sized objects do not fully cover the parameter space, and are inconsistent with observations.
Therefore, shedding light on these processes is key for advancing the knowledge of planet formation.
More specifically, by understanding the details of grain growth during planet formation can shed light into how Earth was formed.

Current research suggests that growth of particles happens mainly in the mid-plane of protoplanetary disks, since the concentration of solid masses in this region is expected to be higher \REF{Testi2014}.
The gas is still coupled to the dust in this phase, and follows the Epstein regime, whereas Brownian motion of micron-sized particles causes slow collisions between particles, leading to grain growth~\REF{Blum1996}.
The interaction between particles can involve different processes, such as sticking (Van Der Waals forces), mass transfer, and fragmentation.
However, these processes are poorly understood.

A widely used approach to simulate grain growth is to develop numerical models (\textit{e.g.} Ref.~\REF{Zsom2010}), or through direct observation using ground- and space-based telescopes.
However, these latter ones are only able to observe static distributions of particle size, and cannot follow different physical processes influencing the formation of large grains and planetesimals.
Nonetheless, several laboratory studies have been conducted to recreate dust aggregation in protoplanetary disks~\REF{Blum2008,Guttler2010,Brisset2015}.

\begin{figure}[!htb]
    \centering
    \includegraphics[width=\dustexperimentsSIZE]{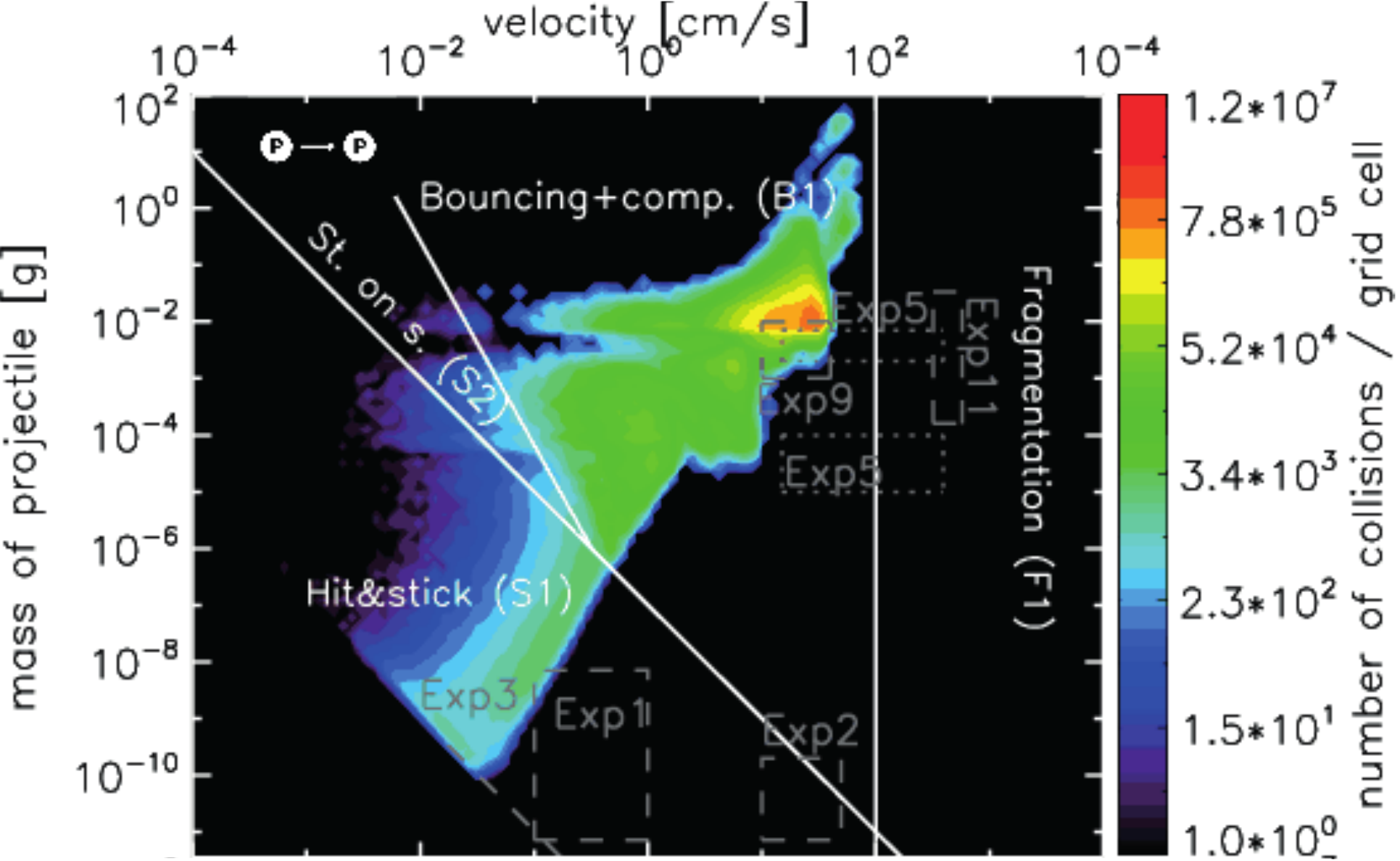} 
    \caption{Parameter space of collision velocity and particle mass, showing number of collisions predicted by a computer simulation of a protoplanetary disk (coloured plot), areas previously studied (dotted boxes), and collision outcomes, such as ``hit \& stick'' and ``fragmentation'' (adapted with permission from~\REF{Zsom2010}).}
    \label{fig:ParticleGrowthExperiments}
\end{figure}

Different properties of the dust grains have been varied to determine the criteria that might lead to dust growth, including size, mass and compositions~\REF{Blum2008}.
An overview of those preceding experiments on dust growth using small porous grains is seen in~\Figure{fig:ParticleGrowthExperiments}.
The majority of collisions do not result in grain growth, which is primarily in the ``hit and stick'' regime (in the lower-left corner of the figure).
However, only a small part of this parameter space has been studied, making it the ideal region to be investigated.

Reproducing the conditions inside a protoplanetary disk on Earth is challenging, since experiments must simulate free-fall conditions, among several other parameters that need to be explored.
So far experiments have been restricted not only to short durations and limited ranges of grain size, but also to a few grain types, with most studies using silicates.
However, to study the influence of dust composition on grain growth, experiments should also use other types of grains, \textit{e.g.} iron rich olivine ones.
Furthermore, water-ice layers on the grains might also have an influence on the grain growth.

Drop tower facilities can provide a micro-gravity environment for up to about \SI{5}{seconds}~\REF{NASA_droptower}.
Although several studies have been performed using drop towers~\REF{Blum2014,Blum2001}, those have used mainly two-particle systems, because of the short experiment duration.

Other experiments have been run on parabolic flights and sounding rockets, which provide around \SIrange[allow-number-unit-breaks=true]{20}{60}{seconds} of micro-gravity, respectively~\REF{Blum2008}.
However, these limit the volume and temperature conditions of the experiments, both of which are necessary to more accurately reproduce the environment in protoplanetary disks.
Additionally, these experiments are unable to precisely control the range of particle velocities.
Again, mostly singular one-on-one collisions are observed, while longer durations are needed to examine the evolution of these particles as an ensemble.

An example of a sounding rocket test was the Suborbital Particle Aggregation and Collision Experiment (SPACE), on the suborbital rocket flight REXUS 12~\REF{Brisset2013}.
This was designed to study the collision behaviour of submillimeter-sized dust aggregates.
Another sounding rocket experiment, the Cosmic Dust Aggregation (CODAG), aimed to investigate the Brownian motion-driven aggregation of cosmic dust~\REF{NASA_sounding}.

An alternative to ground based experiments, is to use the International Space Station (ISS)~\REF{Love2014, Brisset2015}.
Two dust aggregation experiments have been conducted there, and it was found that angular sub-millimetre particles rapidly formed clusters strong enough to survive turbulence in a protoplanetary nebula.
Smaller particles aggregated more strongly and quickly than larger ones, while particle compositions was not found to have a strong effect.
Finally, round, smooth particles aggregated weakly or not at all~\REF{Love2014,Brisset2015}.

Even though the ISS can run long duration experiments, the dust containers had to be shaken every \SI{60}{seconds}, to remove particles stuck to the walls.
There are also significant disturbances to micro-gravity conditions, caused by the astronauts' movement, and a large number of uncontrolled electric and magnetic fields.
These effects make the ISS less suitable, than a dedicated satellite, for conducting experiments with sub-millimetre per second grain collision relative velocities.

Up to now, only one satellite mission is know to have been proposed, the CubeSat Q-PACE~\REF{Colwell2017}.
This CubeSat is under development to explore the fundamental properties of low-velocity ($< \SI{10}{\centi \metre \per \second}$) particle collisions.
It is expected to be launched in first quarter 2018, and ran for three years.

%% file: 2_ScienceObjectives.tex
\section{Scientific objectives}
\label{sec:Science_Objectives}

There are several questions that should be answered, to have a better insight to the mechanisms by which \si{\micro \metre}-sized grains grow in protoplanetary disks.
To fill some of the gaps discussed above, the behaviour of dust grains colliding at low relative velocities ($< \SI{5}{\milli \metre \per \second}$), in micro-gravity ($<\SI{9.8e-6}{N}$), over long time scales (up to several weeks), and with a much larger experimental volume (which allows the dust cloud to evolve), must be observed.
In particular, the target is collisions between particles of sizes $r_{\mathrm{eff}}$ = \SIrange{30}{100}{\micro \metre}, where $r_{\mathrm{eff}}$ is the equivalent radius of a sphere with the same volume as the grain.

Although the ISS and CubeSat experiments can be considered as a baseline for this study, the intention is to go one step further, and to answer the following questions.





\paragraph{How does the size of particles evolve with time, and how does size affect the outcome of collisions?}
To determine grain growth, the sizes of particles before, $r_{\mathrm{eff}_1}$ and $r_{\mathrm{eff}_2}$, and after, $r_{\mathrm{eff}_f}$ , the collisions have to be measure.
Moreover, experiments should be conducted with both mono- and poly-disperse initial grain size distributions.
Mono-disperse distributions should use grains of $r_{\mathrm{eff}} =$ \SIlist{1;30;50;100}{\micro \metre}, whereas poly-disperse ones can be split into two ranges, a wide one ($r_{\mathrm{eff}} =$ \SIrange{0.1}{100}{\micro \metre}) and a narrow range ($r_{\mathrm{eff}} =$ \SIrange{20}{30}{\micro \metre}).
Within both ranges, the particle size distribution should follow a power-law as $n(r_{\text{eff}}) \propto (r_{\mathrm{eff}} / 5 \si{nm})^{-3.5}$~\REF{Pollack1985}.




\paragraph{Does grain shape influence grain growth?}
A random distribution of grain shapes must be used, because even though shapes can be approximated to second order using a continuous distribution of randomly oriented ellipsoids (CDE), this approximation is not realistic.



\paragraph{How does velocity affect grain growth?}
This requires measurements of velocities before collisions, as well as sizes of particles before and after collisions.
Furthermore, measurements should be made in all three axes, and shall provide analysis of the structure of \SI{10}{\micro \metre} grains.


\paragraph{How does rotation of the colliding particles affect grain growth?}
The relative velocity between particles depends not only on the translational one, but on the rotational velocity as well.
For example, if a cluster of dust with a \SI{5}{\micro \metre} size, or larger, rotates with more than \SI{40}{revolutions\,per\,second}, the centripetal force exceeds the Van der Waals force between the dust grains, and fragmentation of the cluster is expected.
Therefore, if clusters rotating at higher frequencies without fragmenting are observed, it is an indication of the presence of additional forces binding the particles together.
To study the effect of the rotation of colliding particles, the angular frequency, axis of rotation and precession should be determined.
The frequencies of interest are between zero and \SI{60}{revolutions\,per\,second}.


\paragraph{How does composition influence the grain growth?}
Silicates have been observed in various parts of protoplanetary disks.
However, the location and physical state of iron in protoplanetary disks still has to be confirmed.
Due to iron depletion in the gas surrounding a young star, it is inferred that iron is present in protoplanetary disk dust particles~\REF{Keller2002}.
Moreover, the effect of ice coatings on grain growth has only been studied to a limited extent.
Various studies show that ice can increase the probability of mass transfer and particle adhesion~\REF{Dominik1997,Wada2007,Wada2008}.
To study these effects, a measurement of the change in particle size as a function of the composition of the colliding grains is required.
Therefore, the proposal is to study SiO$_2$, $\rho_{Si} =$ \SI{2.2}{\gram \per \centi \metre \cubed}~\REF{Rumble2017}, and Fayalite (Fe$_2$SiO$_4$), $\rho_{Fa} =$ \SI{4.4}{\gram \per \centi \metre \cubed}, both with and without a coating of water ice.


\paragraph{How do temperature changes influence cohesion in icy agglomerates?}
Several experiments shall use dust particles coated with an ice layer.
The objective is to measure if the agglomerates fragment or remain whole when ice sublimates, which means measuring particle size before and after sublimation.
To sublimate the ice dust particles must be heated to about \SI{300}{\kelvin} (for one hour).
With this, the experiment mimics what happens to ice-coated agglomerates in a protoplanetary disk, which are heated as they moved towards the central star.

\paragraph{How does porosity affects grain growth, and how do collisions of particles change the porosity?}
Particle porosity may affect the outcome of collisions, with particles with high porosity expected to absorb the impact energy and sticking together, whereas low porosity particles may fragment more easily.
Thus, particles with filling factors, $\phi$, between 0.15 and 0.65, shall be released initially.
During the experiment, either measurements of porosity before and after collisions shall be performed, or the distributions of porosity among the particles before and after the experiments must be obtained.

\begin{figure}[!htb]
  \centering
  \includegraphics[width=\collisiontypesSIZE]{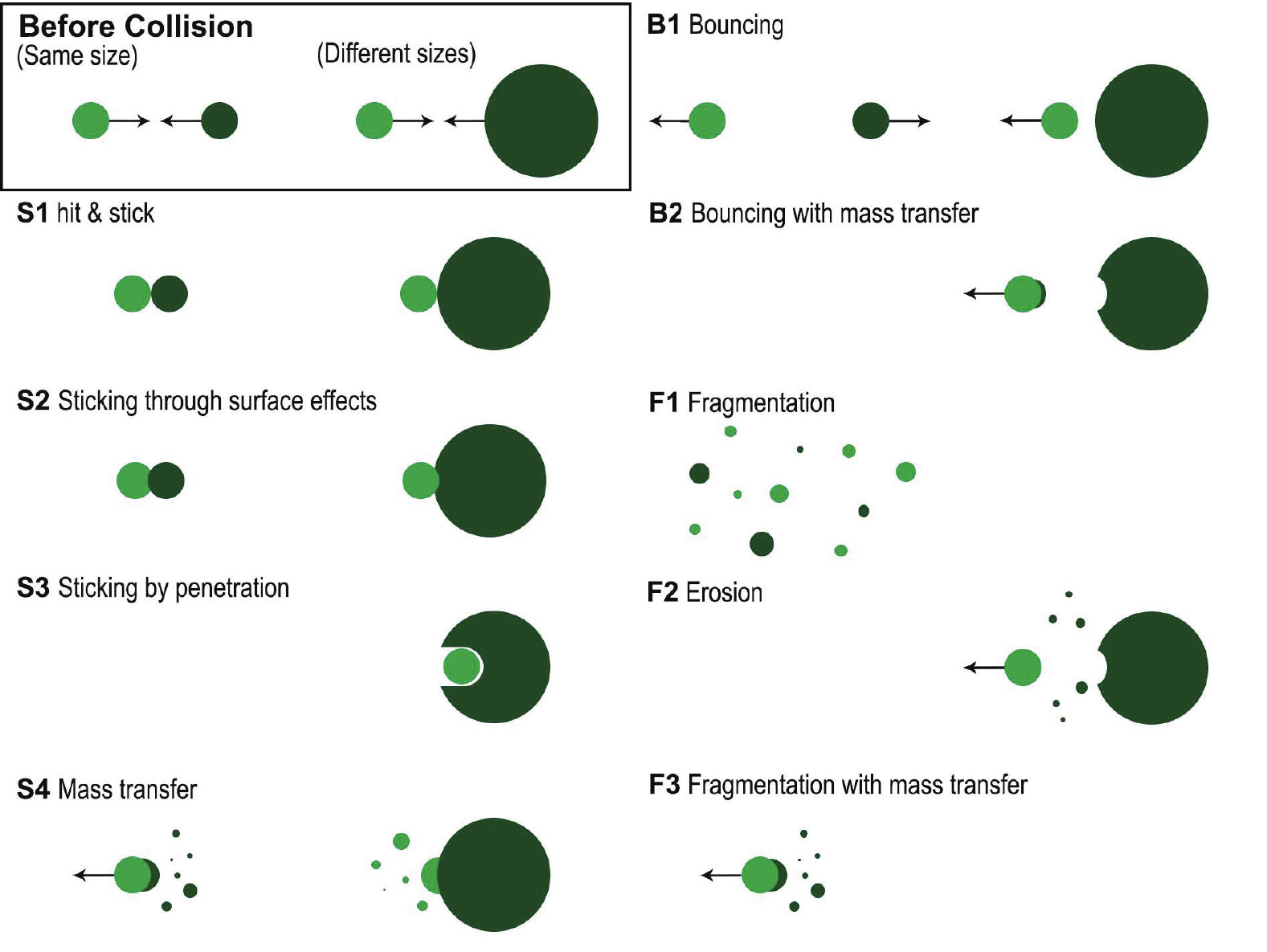} 
  \caption{Classification of different collision outcomes, between particles of the same or distinct sizes. The outcomes are various versions of sticking (S), bouncing (B) and fragmentation (F), adapted from~\REF{Guttler2010}.}
  \label{fig:collision_outcome}
\end{figure}

\paragraph{What are the collisional products?}
The outcome of particle collision can be different variations of sticking, bouncing, and fragmentation.
The same nine different types to classify the collisional outcome of Ref.~\REF{Guttler2010} are used (depicted in~\Figure{fig:collision_outcome}).
%

All of these questions can be translated into a single primary scientific objective (PSO), and several secondary objectives (SSO), as shown in~\Table{tab:Scientific_Objectives_Requirements}.
To fulfil those objectives, maintain a stable environment, and acquire good statistics, several requirements have to be met.
While specific measurement requirements are described in~\Table{tab:Scientific_Objectives_Requirements}, there are some more general ones that apply to all experiments.

\begin{table*}[tbp]
    \scriptsize
    \centering
    \caption{Primary and secondary scientific objectives (PSO and SSO respectively), and corresponding measurement requirements.}
    \label{tab:Scientific_Objectives_Requirements}
    \begin{tabular}{m{8.5cm} m{8.5cm}}
        \toprule
            \textbf{Objectives} 
                & \textbf{Measurement Requirements} \\
        \midrule
            \textbf{\underline{PSO1}:} Measure representative grain size distributions over time, for at least three initially mono-disperse and at least two poly-disperse particle size distributions.
                & \textbf{\underline{PMR1.1}:} Spatially resolve particles and grains in the range $r_{\text{eff}} =$ \SI{0.1}{\micro \metre} to \SI{1}{\milli \metre}.\\
            &\\
            \textbf{SSO1:} Classify collision type as a function of incident grain size.
                & \textbf{SMR1.1:} Classify collision types according to Ref.~\REF{Guttler2010}.\\
            &\\

            \multirow{2}[0]{\hsize}{\textbf{SSO2:} Measure number and sizes of grains after collisions, as function of incident grain sizes (as suggested in Ref.~\REF{Blum2008})}.
                & \textbf{SMR2.1:} Spatially resolve particles and grains as in PMR1.1.\\
                & \textbf{SMR2.2:} Count the number of resulting particles.\\
            &\\
            \multirow{2}[0]{\hsize}{\textbf{SSO3:} Measure the sizes of grains after collisions, as a function of incident grain shape type.}
                & \textbf{SMR3.1:} Spatially resolve particles and grains as in PMR1.1.\\
                & \textbf{SMR3.2:} Reconstruct the shape of the incident particles with a size $r_{\text{eff}} >$ \SI{10}{\micro \metre}.\\
            &\\
            \multirow{4}[0]{\hsize}{\textbf{SSO4:} Measure the change in grain size during collisions, as a function of the frequency of rotation, and relative angle of rotation axes of the incident grains.}
                & \textbf{SMR4.1:} Measure the rotation frequency between \SIrange{0}{60}{\per \second}, to an accuracy of 1\%.\\
                & \textbf{SMR4.2:} Measure the angular velocity vector to an accuracy of 1\%.\\
                & \textbf{SMR4.3:} Measure the precession vector to an accuracy of 1\%.\\
                & \textbf{SMR4.4:} Spatially resolve particles and grains as in PMR1.1.\\
            &\\
            \multirow{2}[0]{\hsize}{\textbf{SSO5:} Measure the change in grain size, as a function of the relative incident velocity of the colliding particles.}
                & \textbf{SMR5.1:} Measure the velocity vector to an accuracy of 1\%.\\
                & \textbf{SMR5.2:} Spatially resolve particles and grains as in PMR1.1.\\
            &\\
        %
        %
            \multirow{2}[0]{\hsize}{\textbf{SSO6:} Measure the change in size of the colliding particles, as a function of the composition of the colliding grains (as suggested in Ref.~\REF{Heiselmann2007})}.
                & \textbf{SMR6.1:} Spatially resolve particles and grains as in PMR1.1.\\
                & \textbf{SMR6.2:} Record the composition of the particles for each experiment.\\
            &\\
            \multirow{2}[0]{\hsize}{\textbf{SSO7:} Measure the change in size of colliding particles, as a function of the initial porosity of the colliding grains.}
                & \textbf{SMR7.1:} Use mono-disperse \SI{30}{\micro \metre} particles, with a filling factor distribution between 0.35 and 0.65.\\
                & \textbf{SMR7.2:} Spatially resolve particles and grains as in PMR1.1.\\
            &\\
            \multirow{3}[0]{\hsize}{\textbf{SSO8:} Measure the influence of ice sublimation on ice agglomerates.}
                & \textbf{SMR8.1:} Spatially resolve particles and grains as in PMR1.1, before and after temperature increase.\\
                & \textbf{SMR8.2:} Measure temperatures between \SIlist{200;300}{\kelvin}, to an accuracy of \SI{20}{\kelvin}.\\
                & \textbf{SMR8.3:} Maintain a temperature above \SI{300}{\kelvin} for at least \SI{1}{\hour}.\\
        \bottomrule
    \end{tabular}
\end{table*}

The first general requirement is related to the dimensions and characteristics of the experimental volume.
Firstly, it has to be large enough so that the particles mean free path ($\lambda_{\text{f}}$) be at least two orders of magnitude larger than the grain size.
This way, it is ensured that the time of interaction between two particles is much shorter than the time between the interactions.
Moreover, with a large volume, the probability of particle to wall collisions is much lower than the probability of particle to particle collisions.
This is important since particle to wall collisions influence the results, and their statistical extrapolation.
Secondly, the number of dust particles that stick to the walls must be minimised, because sticking reduces the amount of particles in the experimental volume, and thus the number of particle to particle collisions.

The second requirement is linked to the use of ice layers.
During the experiments the particle temperature must be low enough to preserve an ice layer for the ice experiments.
Conversely, at the end, when heat is applied, the ice has to be fully sublimated, such that no drops of liquid water are present.
This can be achieved by maintaining the pressure beneath the triple point of water (\textit{i.e.} the pressure in the chamber should be between \SIrange{0.1}{6}{\milli \bar}).
To ensure solid water-ice layers on the grains at a pressure of \SI{0.1}{\milli \bar}, the temperature must be kept beneath 200K.

The magnetic field in the experimental volume also needs to be measured, so to account for any effect it may have on the observed particle behaviour.
Since the main magnetic field is that of the Earth, which varies on the order of minutes, its measurement should occur every \SI{30}{seconds}.

The number of collisions per unit volume and per unit time is given by $N_{\mathrm{coll}} = n_1 n_2 \sigma \delta v$, where $n_1$ and $n_2$ are the number density of particles with effective radii $r_{\mathrm{eff}_1}$ and $r_{\mathrm{eff}_2}$, and $\delta v$ the relative velocity of the colliding particles.
The first-order approximation to the interaction cross-section is given by $\sigma = \pi \left(r_{\mathrm{eff}_1} + r_{\mathrm{eff}_2} \right)^2$.
Considering the computer simulations of Ref.~\REF{Zsom2010}, at least 10$^6$ collisions have to be observed to also cover unlikely collision events (\textit{e.g.} collisions between particles with a size of $r_{\text{eff}} >$ \SI{1}{\milli \metre}).
At the same time, the condition $\lambda_{\text{f}} \gg r_{\text{eff}}$ must be met.

An overview of the different types of experiments, with composition, porosity, size, and distribution, can be found in~\Table{tab:Experiments}.
\begin{table}[htbp]
    \footnotesize
    \centering
    \caption{Characteristics of all 28 experiments, performed at a temperature of \SI{200}{\kelvin}, pressure range of \SIrange{0.1}{6}{\milli \bar}, and with velocities between \SI{1}{\micro \metre \per \second} and \SI{5}{\milli \metre \per \second}.}
    \label{tab:Experiments}
    \begin{tabular}{>{\centering}m{2cm} c c m{2cm}<{\centering}}
    \toprule
        \textbf{Grain size distribution}    & \textbf{Grain size [\si{\micro}]} & \textbf{Composition}  & \textbf{Porosity Distribution} \\
    \midrule
        Mono                                & 1, 30, 50, 100                    & SiO$_2$               & 0.15-0.35 \\
        Mono                                & 1, 30, 50, 100                    & Fayalite              & 0.15-0.35 \\
        Mono                                & 1, 30, 50, 100                    & SiO$_2$ + Ice         & 0.15-0.35 \\
        Mono                                & 1, 30, 50, 100                    & Fayalite + Ice        & 0.15-0.35 \\
        Mono                                & 30                                & SiO$_2$               & 0.25-0.35 \\
        Mono                                & 30                                & SiO$_2$               & 0.35-0.45 \\
        Mono                                & 30                                & SiO$_2$               & 0.45-0.55 \\
        Mono                                & 30                                & SiO$_2$               & 0.55-0.65 \\
        Poly                                & 20-30, 0.1-100                    & SiO$_2$               & 0.15-0.35 \\
        Poly                                & 20-30, 0.1-100                    & Fayalite              & 0.15-0.35 \\
        Poly                                & 20-30, 0.1-100                    & SiO$_2$ + Ice         & 0.15-0.35 \\
        Poly                                & 20-30, 0.1-100                    & Fayalite + Ice        & 0.15-0.35 \\
    \bottomrule
    \end{tabular}
\end{table}

%% file: 4_MissionDesign.tex
\section{Mission design}
\label{sec:Mission_Design}

\subsection{Mission objectives and drivers}
\label{sec:Mission_Objectives}

Considering the scientific objectives and requirements, described in~\RSection{sec:Science_Objectives}, the two main mission requirements are to conduct all planned science experiments (listed in~\Table{tab:Experiments}), and to retrieve the data products generated by the payload (see~\RSection{sec:Payload_Description}).
Therefore, the approach here is to put in space a laboratory, Magrathea, and such provide a continuously undisturbed micro-gravity environment, where several experiment parameters can be monitored and controlled, allowing observations of grain growth over three orders of magnitudes in size, from the \si{\micro \metre} to the \si{\milli \metre} scale.

The four main drivers that have been identified for this mission are:
\begin{enumerate}
    \item The continuous micro-gravity environment, since the experiments need to be protected from major external disturbances;
    \item The time to complete all experiments, including the time taken to clean the chamber from contaminants between experiments;
    \item The control of the experimental volume environment, so that pressure and temperature remain homogeneous over the whole volume;
    \item The amount of data generated by the experiments, for which the service module must provide sufficient downlink capability to effectively transfer this data to Earth.
\end{enumerate}

\subsection{Launcher, orbit and environment}
\label{sec:Launcher_Orbit}

The satellite was assumed to be placed into its initial orbit by a Soyuz vehicle, launched from Guiana Space Centre.
This is a high reliable and accurate launcher, capable of an orbit altitude uncertainty of $\pm$ \SI{12}{\kilo \metre}, and just $\pm$ \ang{0.12} for the inclination, with a $3\sigma$ confidence level~\REF{Arianespace2012}.

The final operational orbit is selected to be a \SI{800}{\kilo \metre} Sun-Synchronous Orbit (SSO), with an inclination of \ang{98.6}, and a Local Time of Ascending/Descending Node of 0600/1800.
While at low Earth orbit (LEO) aerodynamic drag is significant, at higher orbital altitudes the radiation intensity from the Van Allen belt increases.
Therefore, the \SI{800}{\kilo \metre} orbit was selected so that the Van Allen belt disturbances are manageable, and the drag force is lower and within the same order of magnitude as the solar radiation pressure (about \SI{1e-7}{\metre \per \second \squared}).
Moreover, with a SSO orbit there are long access times for downlink, increased thermal stability due to the constant sun angle, and no attitude manoeuvres are required for solar array pointing.

To monitor the effect of the Earth's magnetic field, the payload includes a magnetic field measurement system (see~\RSection{sec:Payload_ExperimentalVolume}).
To mitigate the risk of leaving any debris on the final orbit, an initial orbit altitude of \SI{785}{\kilo \metre} is targeted, followed by an orbit correction manoeuvre.


At the spacecraft End Of Life (EOL), and since its mass is expected to be above \SI{1000}{\kilogram}, a de-orbiting manoeuvre is planned to ensure safe disposal.
Assuming a standard disposal time of \SI{25}{years}, the perigee is lowered to \SI{520}{\kilo \metre}.
From that point forward, the atmospheric drag is sufficient to dispose of the spacecraft in the required time.

The $\Delta v$ budget for the full mission, including station keeping and collision avoidance, is shown in~\Table{tab:DeltaV Budget}.
\begin{table}[!htbp]
    \footnotesize
    \centering
    \caption{Total $\Delta v$ budget analysis, where station keeping is assumed to be required for the mission lifetime of \SI{5}{years}.}
    \label{tab:DeltaV Budget}
    \begin{tabular}{ccc}
        \toprule
            \textbf{Manoeuvre}                  & \textbf{$\Delta v$ [\si{\metre \per \second}]}    & \textbf{Fuel mass [\si{\kilogram}]} \\
        \midrule
            Injection                           & 38                                                & 15 \\
            Station-keeping (inclination)       & 25.5                                              & 8.5 \\
            Station-keeping (drag compensation) & 12                                                & 4 \\
            Collision avoidance                 & 0.5                                               & 0.2 \\
            De-orbiting                         & 75                                                & 31 \\[0.8ex]
            \textbf{Total}                      & \textbf{151}                                      & \textbf{58.7} \\
        \bottomrule
    \end{tabular}
\end{table}

\subsection{Ground Segment}
\label{sec:Ground_Segment}

For satellites in near-polar orbits, a single high-latitude ground station can provide good coverage.
The KIR-1 \SI{15}{\metre} antenna at Kiruna Estrack ground station, in northern Sweden, is a good option for Magrathea.
If the KIR-1 antenna is unavailable, KIR-2 antenna can be used as a backup, so that communications are not affected.

With these antennae, and the spacecraft in a \SI{800}{km} SSO orbit, there will be 12 access windows per day, with a mean duration of \SI{730}{\second} per window.
Most of the time, the antenna will receive science data from the payload, since housekeeping only requires a few \si{kbps} (see~\RSection{sec:Spacecraft_Communications}).

\subsection{Mission timeline}
\label{sec:Mission_Timeline}

The nominal mission orbital lifetime is about five years, including commissioning time, as listed in~\Table{tab:OverallConOps}.

\begin{table}[htbp]
    \footnotesize
    \setlength{\tabcolsep}{2pt}
    \centering
    \caption{Concept of operations of the Magrathea mission (for details on experiment operation see~\RSection{sec:Payload_ExperimentProcedure}).}
    \label{tab:OverallConOps}
    \begin{tabular}{c >{\centering}m{4.5cm} c}
        \toprule
            \textbf{Phase}                      & \textbf{Details}                                  & \textbf{Length} \\
        \midrule
            \multirow{2}[0]{*}{Pre-flight}      & Assembly test and launch operations               & 9 months \\
                                                & Mount spacecraft on launcher                      & 2 months \\
            & & \\
            \multirow{3}[0]{*}{Launch}          & Soyuz launcher                                    & \multirow{3}[0]{*}{\SI{1}{day}} \\
                                                & Sun-Synchronous \SI{800}{\kilo \metre} orbit      &  \\
                                                & Shared ride if possible (mass < \SI{2}{\tonne})   &  \\
            & & \\
            \multirow{2}[0]{*}{Detumble}        & Stabilize                                         & \multirow{2}[0]{*}{\SI{1}{day}} \\
                                                & Acquire attitude                                  &  \\
            & & \\
            \multirow{3}[0]{*}{Commissioning}   & Deploy antennas and solar arrays                  & \multirow{3}[0]{*}{\SI{30}{days}} \\
                                                & System check                                      &  \\
                                                & Start science operations                          &  \\
            & & \\
            \multirow{4}[0]{*}{Experiment}      & Release sample                                    & \multirow{3}[0]{*}{\SI{5}{years}} \\
                                                & Take continuous measurements                      &  \\
                                                & Take close-up measurements                        &  \\
                                                & Vent chamber                                      &  \\
            & & \\
            \multirow{2}[0]{*}{End of lifetime} & De-orbit burn                                     & \multirow{2}[0]{*}{Up to \SI{25}{years}} \\
                                                & Final perigee of \SI{520}{\kilo \metre}           &  \\
        \bottomrule
    \end{tabular}
\end{table}

If the spacecraft is still operating according to the science requirements, and there is interest to keep the last experiments running for longer periods, a mission extension can be investigated.
This extension would most likely be contained within one year, since at this time any other situation that would justify such extension is not foreseen.


%% file: 3_Payload.tex
%
%
%

\section{Payload}
\label{sec:Payload_Description}

\subsection{Requirements}
\label{sec:Payload_DrivingRequirements}

The science objectives have been translated to measurement requirements (\Table{tab:Scientific_Objectives_Requirements}), which specify the range and precision of measurements for each variable under study.
These variables include dust size, shape, composition, porosity, presence or lack of an ice layer, collision velocity and type, frequency of rotation, and relative angle of rotation axes.
The measurement requirements are then used for science instrument selection.

Apart from measurement requirements, there are also requirements on sample handling and processing, and on the experimental volume.
These requirements concern the duration of each experiment, sampling rate, and cleanliness, as follows.

The duration of each experiment is set by the need to observe at least $10^6$ collisions, whilst maintaining a dust grain mean free path below 0.01 of the smallest dimension of the experimental volume.
This ensures that particle-particle collisions dominate particle-wall collisions.

To effectively monitor the variation of physical quantities such as size, the experimental volume should be sampled for dust grain contact measurements at least every two hours.

Requirements on cleanliness have been defined to address the risk of contamination between experiments.
In particular, no more than 20\% of particles may stick to the walls of the experimental volume at any time.
Furthermore, no more than 1\% may remain inside the experimental volume after each experiment.
At the same time, any agitation manoeuvre used to free particles from the walls of the experimental volume must not result in particle speeds above \SI{2}{\milli \metre \per \second}, so that the low-velocity collision regime is preserved.

\subsection{General description}
\label{sec:Payload_GeneralDescription}

Science instruments are divided into two categories: far-range instruments, which observe particles at a distance, and close-range instruments, which perform measurements on a subset of particles extracted from the experimental volume.

The far-range instruments include three particle tracking cameras (P-CAMs), which measure particles from middle size to the biggest ones.
The close-range instruments are an Optical Microscope (OM) and an Atomic Force Microscope (AFM), which cover the lower portion of the particle size range.
The estimated mass, power, data rate, and volume of these instruments are given in~\Table{tab:Instruments}.

\begin{table}[!htbp]
    \footnotesize
    \setlength{\tabcolsep}{2pt}
    \centering
    \caption{Characteristics of the three science instruments.}
    \label{tab:Instruments}
    \begin{tabular}{l >{\centering}m{2cm} >{\centering}m{2cm} m{2.3cm}<{\centering}}
    \toprule
                                & Particle tracking camera (P-CAM)  & Optical Microscope (OM)   & Atomic Force Microscope (AFM) \\
    \midrule
        Mass [\si{kg}]          & 1.5                               & 1.1                       & 8.3 \\
        Power [\si{W}]          & 4                                 & 1                         & 17 \\
        Data rate [\si{Mbps}]   & 5.4                               & 0.075                     & 0.001 \\
        Dimensions [\si{cm}]    & 7 $\times$ 7 $\times$ 7           & 7 $\times$ 5 $\times$ 9   & 30 $\times$ 25 $\times$ 10 \\
    \bottomrule
    \end{tabular}
\end{table}

There are two other major payload elements, the experimental volume, in which particle collisions take place, and the Sample Handling and Processing (SHP) subsystem.
This latter one carries out all functions related to sample storage, injection, extraction, transportation, and venting.

\begin{figure*}[htbp]
    \centering
    \includegraphics[width=\payloadoverviewSIZE]{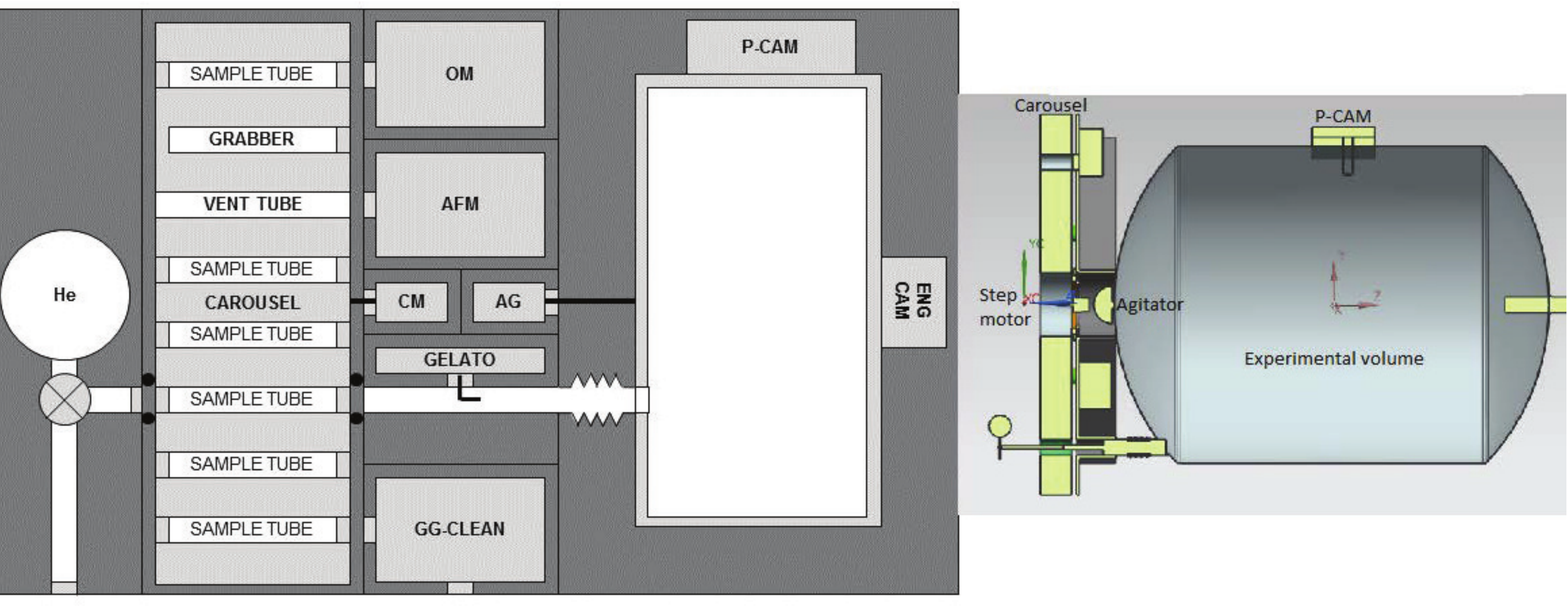} 
    \caption{Schematic (left) and CAD model (right) of the payload (see text for explanation).}
    \label{fig:payload_layout}
\end{figure*}

A schematic and CAD model of the payload is shown in~\Figure{fig:payload_layout}.
At the far left, the payload includes a helium tank for dust fluidisation, a master valve, and an overboard vent path.
The vent path is connected to the carousel, which contains the sample tubes, the Grain Grabber (GG), and a vent tube.
The next payload section houses the close-range instrumentation (OM, AFM), Carousel Motor (CM), the experimental volume agitator (AG), the injection path (including GELATO for ice layer generation), and the Grain Grabber cleaning system (GG-CLEAN), with its associated overboard vent path.
Finally, on the far right, the experimental volume is shown with its attached particle tracking cameras (P-CAMs) and engineering camera (ENG CAM).

\subsection{Far-range instruments}
\label{sec:Payload_FarRangeInstruments}

The measurement of particle positions and linear and angular velocities in three dimensions requires at least two cameras.
Since observation of collisions is fundamental to mission success, a third camera is added for redundancy and improved precision.
The three cameras, aligned orthogonally, observe a \SI{1}{\cubic \centi \metre} volume with a spatial resolution of \SI{3}{\micro \metre} and a frame rate of \SI{120}{fps}.
Each camera has a focal plane array of \SI{3400 x 3400}{pixels}.

Due to the high data rate for each camera (approximately \SI{14}{\giga \byte \per \second}), substantial on-board compression is required.
On-board software processes the images to extract the required numerical data (particle positions and velocities), reducing the data volume to a manageable amount for downlink.
However, some images are needed to assess particle shapes and collision types.
Therefore, ``snapshots'' of individual particles, with effective radii above \SI{15}{\micro \metre}, are taken and downlinked in a compressed image format.
With these two types of on-board processing, a data rate of approximately \SI{17}{\mega \byte \per \second} is achieved.





\subsection{Close-range instruments}
\label{sec:Payload_CloseRange Instruments}

Since far-range instruments can only resolve particles with sizes greater than \SI{3}{\micro \metre}, additional instrumentation is required to meet the \SI{0.1}{\micro \metre} minimum particle size measurement requirement.
Moreover, properties such as porosity and shape of moving particles are extremely difficult to measure remotely.
Therefore, the plan is to periodically extract a subset of grains from the experimental volume, such that they can be presented to the OM and AFM for close-range analysis.

The OM can be used to assess the size, shape, and porosity of particles in the mid size range.
It can also be a diagnostic tool to verify that the grain grabber, the device used to extract grains from the chamber, is clean enough before re-insertion.
Optical microscopes have extensive heritage and low technological risk.
A similar design to Rosetta's CIVA-M/V microscope (Ref.~\REF{Bibring2007}) is followed, with an expected diffraction limit at approximately \SI{1}{\micro \metre}.
Below this size another measurement technique is required.


After inspection by the OM, dust grains are measured using the AFM.
This provides size and shape measurement capabilities down to the smallest size required.
The principle behind this involves moving a microcantilever with a sharp tip across a dust grain restrained on a substrate.
Variations in the height of the dust grain's surface cause vertical deflections of the cantilever, detectable by a laser.
Since the deflections are small, AFM allows for extremely high resolution, down to the \SI{10}{\nano \metre} range.
The Phoenix lander and the Rosetta's Micro-Imaging Dust Analysis System (MIDAS) have already used AFMs~\REF{riedler2007midas}.
This flight heritage means that the technology is already well-developed and, although it has higher complexity than the other instruments, remains at a relatively low risk.





\subsection{Experimental volume}
\label{sec:Payload_ExperimentalVolume}

The experimental volume is an aluminium cylinder, whose dimensions are driven by two opposing requirements.
The volume must be small enough such that the collision rate is high enough to reach the required number of collisions in a reasonable amount of time, but large enough such that the particle mean free path remains large relative to particle size, to avoid unrepresentative conditions.
Considering these requirements, and the size of typical launcher fairings, a \SI{6}{\cubic \metre} volume was selected.

The experimental volume is mechanically decoupled from the spacecraft structure so that it can be periodically agitated, to remove particles stuck to its inner walls.
This is done using an off-axis motor and a low-amplitude resonator.
Locks secure secure it in place for launch.

Environmental conditions inside the volume must be monitored, in particular pressure, temperature, and magnetic field strength.
The instrumentation to perform this is relatively common (\textit{e.g.} Ref.~\cite{chobotov1993,neubauer1993}), with flight heritage hardware available and assessed as low risk.

\subsection{Sample Handling and Processing (SHP) subsystem}
\label{sec:Payload_SampleHandling}

The Sample Handling and Processing (SHP) subsystem  performs sample storage, injection, extraction, transportation, and venting.
The central part of the SHP is a carousel, which rotates various components to align them with an injection tube connected to the experimental volume.
These components include the individual grain canisters for each experiment, the grain grabber mechanism (described below), and a through hole to allow for chamber venting.
Due to its high mechanical complexity, strict cleanliness requirements, and relative novelty, the SHP subsystem is assessed to be at Technology Readiness Level (TRL) 2, and is likely to pace the development schedule.

\subsubsection{Grain Grabber (GG)}
\label{sec:Payload_GrainGrabber}

The Grain Grabber (GG) retrieves grains from the experimental volume and, by rotating the carousel, presents them to the close-range instruments.
The design comprises a collection plate, to which a charge can be applied to attract and retain grains, and a telescopic linear actuator, which permits the plate to be extended into the experimental volume and retracted into its housing in the carousel.
Once the close-range measurements have been completed, the GG is rotated into alignment with a cleaning system.
This uses piezoelectric plate agitation, gas burst, and venting to vacuum to remove grains.
Although some elements of the GG are common to sample handling mechanisms on other spacecraft, the requirements for this mission are unique and therefore the GG is assessed as being at TRL 2.

\subsubsection{Ice layer generation capability (GELATO)}
\label{sec:Payload_Gelato}

The capability to coat grains with an ice layer is required to fulfil some of the scientific requirements.
The current concept is to disperse water into the empty experimental volume followed by dust grains.
As the dust grains are cooler than the water, the water will condense onto the dust and freeze.
The mechanical design consists of an injector, with a nozzle and a set of swirling vanes, which generates the turbulence necessary to effectively disperse the water.
The thickness of the ice layer is determined by the ratio of total mass of injected water to total dust grain surface area, and will be calibrated during development.
The GELATO is considered to be at TRL 2.

\subsection{Experimental procedure}
\label{sec:Payload_ExperimentProcedure}

A single experiment can be divided into three main phases: preparation, collision (during which sampling takes place), and cleaning.

An experiment begins by aligning the desired grain container in the carousel with the injection port.
An engineering camera, located at the top of the experimental volume, is turned on, and water is dispersed if an ice layer is required.
High-pressure helium is then used to drive the dust grains into the experimental volume.
The injector generates enough turbulence to effectively break up any grain agglomerates that may have formed in storage.

During the collision phase, the dust cloud is left to evolve, as particle tracking cameras record the collisions.
To prevent the excessive particle build-up on the walls, the chamber is periodically agitated.
At set intervals, close-range measurements are performed, by first aligning the GG with the injection port.
The GG is then extended into the volume to collect a sample, before retracting back into the carousel and being aligned with the OM and AFM in turn.
After the close-range observations, the carousel rotates again to present the GG to the cleaning system.
This process can be repeated as required, with someOM inspections to ensure GG cleanliness.

Once the experiment finishes, the experimental volume must be cleaned.
First, the carousel is rotated to allow high-pressure helium to be blown to the experimental volume, and then the volume is agitated.
Finally, the volume is vented to space, with the flow modulated to entrain particles effectively.
This pressurisation, agitation, and venting can be repeated as many times as needed.
Once all valves are closed, the experiment is complete and the process can begin again.

\subsection{Technology development}
\label{sec:Payload_TechnologyDevelopment}

Due to mission novelty, and low instrument maturity, in particular of the grain grabber and the GELATO (currently at TRL 2), a robust technology development programme is required.
Moreover, verification of individual components of the SHP subsystem should first be performed in a laboratory environment, before an integrated sub-system test on parabolic flights.


%% file: 5_Spacecraft.tex
%
%
%

\section{Spacecraft}
\label{sec:Spacecraft}

\subsection{Structure \& mechanisms}
\label{sec:Spacecraft_Structure}

\begin{figure}[!htb]
    \centering
    \includegraphics[width=\MagratheaSIZE]{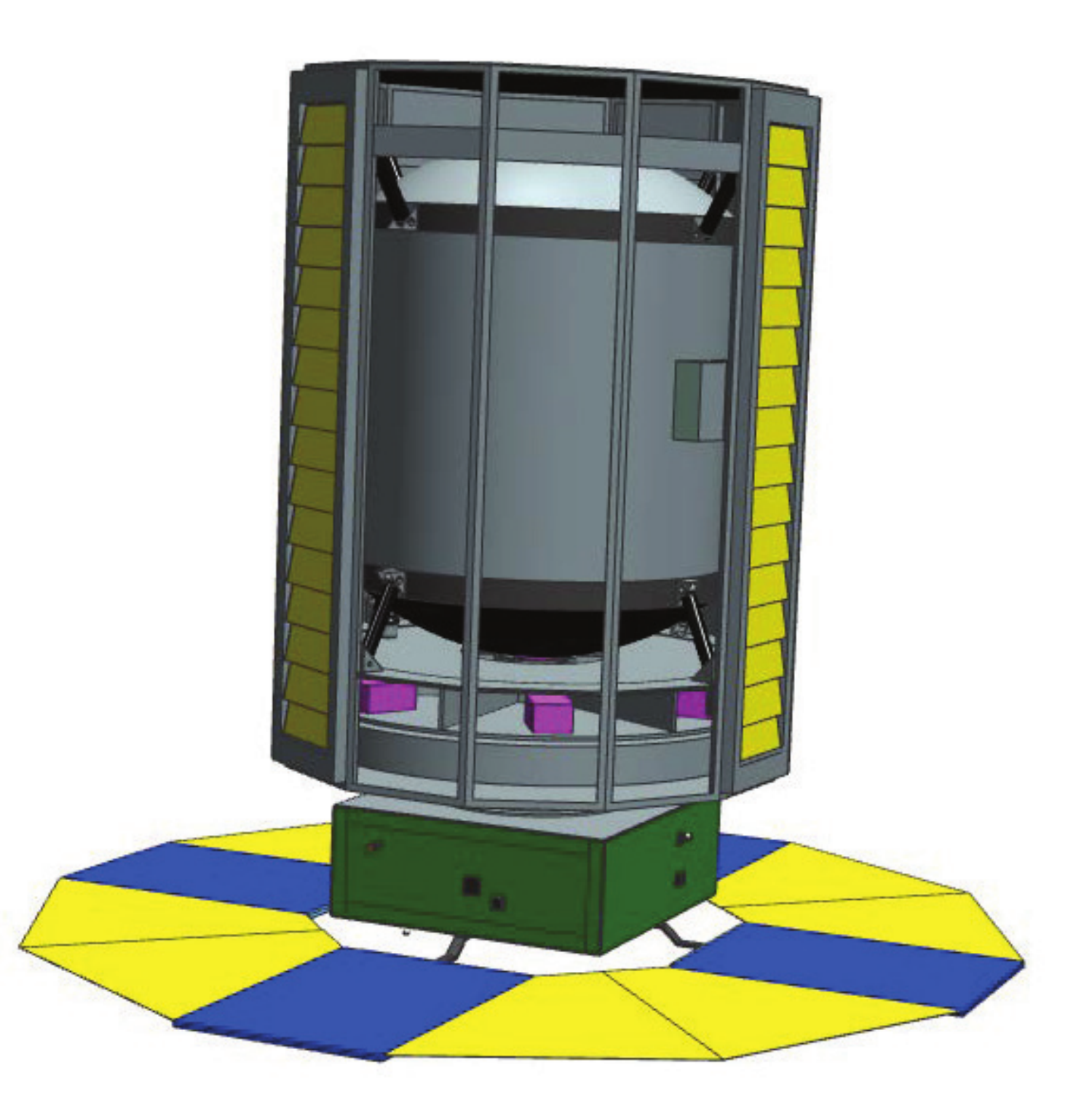} 
    \caption{A view of the spacecraft, showing the experimental volume at top, louvres on the sides, and the service module at the bottom}
    \label{fig:spacecraft_overview}
\end{figure}

After some studies, a model of the spacecraft selected is represented in~\Figure{fig:spacecraft_overview} (in flight mode).
This configuration reflects the driving requirements of the spacecraft structure, namely:
\begin{enumerate}
    \item The size and shape must be compatible with a Soyuz launcher fairing;
    \item It must be capable of withstanding launch loads;
    \item It must accommodate a \SI{6}{\cubic \metre} experimental volume;
    \item It must support deployable solar panels and a sun-shield;
    \item It must support three-axis stabilisation.
\end{enumerate}
%
%

The service module is designed iw a box configuration, with wall panels mounted to the frame for easy access to the interior.
The instruments placement inside was optimised to align the bus centre of mass with the spacecraft's axis.
The solar panels and sun-shield are mounted to one side of the bus platform, with the use of an industry-standard solar array drive mechanism.
These fold when the spacecraft is in the stowed configuration.

Launch loads are carried by a central column, manufactured from a filament-wound Carbon Fibre-Reinforced Plastic (CFRP).
The main propellant tank is mounted inside this central column and covered with thermal insulation.
Aluminium is used for other primary load-bearing structures and fittings, while secondary low loading structures are made of CFRP honeycomb-core panels.

The payload, comprising the experimental volume, sample handling and processing subsystem, and science instruments, is mounted inside a 12-sided frame.
This is connected to the central cylinder by a flange interface.
The experimental volume is mounted on sealed springs to allow for agitation, while avoiding excessive vibration during launch.
The sample handling and processing subsystem and close-range science instruments are mounted on shelves which are connected directly to the frame.

\subsection{Thermal control}
\label{sec:Spacecraft_Thermal}

The main heat fluxes during a full orbit are the Sun's radiation, the infrared radiation (IR) from the Earth, and the Sun radiation reflected by the Earth.
For a safe operation of the electronics and batteries, the allowed temperature range within the bus is from \SIrange{255}{320}{\kelvin}, whereas the payload structure needs to maintain a temperature below \SI{200}{\kelvin}.
Therefore, the thermal subsystem is not only complex, but it also needs active control.

When the spacecraft sees the Sun, it should keep the solar panels (and thus the sun-shield) always pointing at the Sun.
This way, the experimental volume is protected from direct sunlight.
Because the back of the spacecraft is facing the Earth's horizon, it is coated with a quartz mirror.

The bus must be covered with a Multi Layer Insulation (MLI), to protect it from rising temperatures.
Additionally, heaters are used during eclipse, to maintain the temperatures within the predefined range.
So the spacecraft health can be monitored, and to control heater operations, the temperatures are measured using thermocouples.

To minimise heat exchange between the bus and the payload, the surface of the latter facing the bus is also covered with MLI.
Furthermore, the bus and the payload are held together by a low conducting structure.

Temperature variations of the experiment chamber, due to heat gradients, are mitigated by using a highly conductive material on the chamber itself.
Apart from making the temperature uniform, the surface of the experiment chamber is painted black to maximise radiated heat flow.
Nevertheless, the IR radiation and the reflected sunlight can still warm the payload too much, and thus the structure is covered with louvres.
The louvres facing the Earth are closed, reflecting the radiation (since they are externally coated with aluminium), while the ones facing deep space are opened, emitting radiation.
To reduce the exchanged radiation between the louvres and the internal structure, these are coated with vaporised deposited gold on the inside.

A simplified and steady-state model was used to calculate the equilibrium temperatures for the payload structure and bus during both solar illumination and eclipse, shown in~\Table{tab:exp_chamber}.
The resulting temperatures for the payload, with the use of louvres, are below the upper limit of \SI{200}{\kelvin}, validating the configuration selected.

There is one instance when the louvre operation has to be reversed, and all louvres facing the Earth are opened and the ones facing deep space are closed.
This happens at the end of experiments evolving ice, when the experimental volume has to be heated to \SI{300}{\kelvin}.

\begin{table}[!htbp]
    \footnotesize
    \centering
    \caption{Steady-state approximate temperatures under solar illumination (hot case) and eclipse (cold case) of the payload and bus structure.}
    \label{tab:exp_chamber}
    \begin{tabular}{ccc}
    \toprule
        \multirow{2}[0]{*}{\textbf{Structure}}  & \multicolumn{2}{c}{\textbf{Case temperature [\si{K}]}} \\
    \cmidrule(r){2-3}
                                                & Cold      & Hot \\
    \midrule
        {Payload}                               & 160       & 190 \\
                                                
        {Bus}                                   & 263       & 293 \\
                                                
    \bottomrule
    \end{tabular}
\end{table}

\subsection{Attitude and orbit control}
\label{sec:Spacecraft_Attitude}

The main driver for the Attitude and Orbit Control System (AOCS) is the minimisation of accelerations imparted to particles by the walls of the experimental volume.
Translating it to a subsystem requirement, it means that an acceleration higher than \SI{8e-10}{\metre \per \second \squared} has to be detect, whilst the experiments are being conducted.
Due to the spacecraft high frontal surface area, with the sun-shield and solar panels, the solar radiation pressure will create an acceleration of the order of \SI{1e-7}{\metre \per \second \squared}.
The albedo radiation and atmospheric drag produce accelerations an order of magnitude lower to the side surface.
Thus, the resulting force on the spacecraft is about \SI{200}{\micro \newton}.

With accelerations three orders of magnitude higher than the requirement, a system capable of detecting and counteracting these accelerations had to be devised.
The most important components of the system are a drag-free accelerometer and some Field-Emission Electric Propulsion (FEEP) thrusters.
The accelerometer used on missions like the BepiColombo, LISA Pathfinder, and GOCE missions, has higher accuracies than the one required~\REF{Iafolla2010}.
The FEEP is an AMR nano thruster capable of \SI{10}{\micro \newton}  up to \SI{0.5}{\milli \newton}~\REF{AMRPropulsionTechnologies2017}.

For the rest of the spacecraft operations, when no experiments are being conducted, the AOCS is composed of standard components, as there are no stringent pointing requirements.
To sense the attitude the equipment includes four sun sensors (with $\le \ang{0.1}$ RMS accuracy), two star trackers (with \SI{5}{\arcsecond} cross-boresight), and an inertial measurement unit (with \SI{300}{\micro g} and \SI[number-unit-product = {}]{1}{\degree \per \hour} acceleration and rotation accuracies, respectively)~\REF{NorthropGrumman2017,SinclairInterplanetary2016,NewSpaceSystems2016}.
For attitude control the spacecraft has momentum wheels capable of generating a torque of \SI{45}{\newton \metre} and magnetotorquers generating a maximum dipole momentum of \SI{100}{\ampere \metre \squared}~\REF{AirbusDefense&Space2016,NewSpaceSystems2016a}.

The orbit is determined using a Surrey GNSS, with an error of \SI{10}{\metre} in position, and \SI{0.15}{\metre \per \second} in velocity~\REF{SurreySatelliteTechnology2016}.

\subsection{Power}
\label{sec:Spacecraft_Power}

For LEO and SSO orbits, solar power is the most efficient method, and the one assumed for Magrathea.
Therefore, the power supply chain is composed of solar panels, a Power Distribution and Control Unit (PDCU) and batteries.

Assuming a solar array efficiency of 0.2 and an EOL degradation factor of 0.8, a solar panel area of \SI{6.3}{\square \metre} is needed.
The solar cells are made of GaInP2/GaAs/Ge, which has significant flight heritage~\REF{SpectrolabInc.2010}.

The generated power is managed by the PDCU, with a peak power of \SI{900}{\watt}.
The board produced by ASP is a suitable one, providing triple redundancy~\REF{ASP-EquipmentGmbH2017}.

After orbital propagation, it was seen that a peak shadow phase of \SI{18}{minutes} occurs twice a year.
Thus, a battery subsystem, capable of providing \SI{1000}{\watt \hour}, is needed.
This value is designed to allow a depth of discharge (DOD) of 55\%, with a remaining capacity at EOL of 80\%.
This can be obtained by using ten EaglePicher SLC-028-01 Li-ion cells~\REF{EaglePicherTechnologiesLLC2011}.

\subsection{Communications}
\label{sec:Spacecraft_Communications}

The communication subsystem is driven by the payload data rate requirement.
For sizing the communication subsystem link budgets for Telemetry, Tracking \& Commanding (TT\&C) and payload data downlink were created (shown in~\Table{tab:TT_budget}).

\begin{table}[!htbp]
    \footnotesize
    \setlength{\tabcolsep}{2pt}
    \centering
    \caption{Link budget for TT\&C (S-Band) and payload data downlink (X-Band).}
    \label{tab:TT_budget}
    \begin{tabular}{l >{\centering}m{1.8cm} >{\centering}m{1.8cm} m{1.8cm}<{\centering}}
        \toprule
        \textbf{Characteristic}     & \textbf{S-band uplink (\SI{100}{kbps})}   & \textbf{S-band downlink (\SI{2}{Mbps})}   & \textbf{X-band downlink (\SI{175}{Mbps})} \\
        \midrule
        EIRP [\si{dBW}]             & 32                                        & -2                                        & 18 \\
        Path losses [\si{dB}]       & 157                                       & 157                                       & 169 \\
        G/T [\si{dB/K}]             & -22                                       & 9.6                                       & 37 \\
        EB/N0 [\si{dB}]             & 30                                        & 14                                        & 30 \\
        Required EB/N0 [\si{dB}]    & 8                                         & 8                                         & 14 \\
        Margin [\si{dB}]            & 22                                        & 6                                         & 16 \\
        \bottomrule
    \end{tabular}
\end{table}

Although TT\&C has a low data volume, it is crucial for housekeeping data.
A relatively small data rate of \SI{100}{kbps} for uplink, and \SI{2}{Mbps} for downlink, is sufficient to provide a good energy per bit to noise power spectral density ratio (EB/N0), even when the signal is weak.
Using a modulation scheme such as QPSK with Reed-Solomon FEC (Forward Error Correction), a bit error rate of $10^{-4}$ at most would be obtained.
Therefore, two S-band patch antennas are used, pointing in different directions and with one transceiver per antenna~\REF{EnduroSat2016,Elta2016}.

A specific system for payload data downlink is necessary, as the instruments can generate around \SI{190}{\giga \byte} per day.
One X-band horn antenna can be used, with a data rate of up to \SI{175}{Mbps}~\REF{A-Info2016}, to downlink all data in one day (assuming 12 passes per day).
A modulation scheme, such as 8-PSK with Reed-Solomon FEC, is needed, requiring an EB/N0 ratio of \SI{14}{dB}, to have a bit error rate of $10^{-6}$ or less.

\subsection{On-Board computer \& Data Handling}
\label{sec:Spacecraft_Computer}

The high data volume and the need for on-board image processing drives the spacecraft computer selection.
To comply with both, two On-Board computers and Data Handling (OBDH) subsystems are used.
While the first OBDH subsystem purpose is to process data from the payload and store part in memory, the second one is responsible for acquiring, formatting, and encoding spacecraft telemetry, and data downlink.

For the payload a Sirius C\&DH computer has been selected, which has a 32-bit OpenRISC fault-tolerant processor, a mass memory storage of \SI{16}{\giga \byte}, and a power supply from \SI{4.5}{\volt} to \SI{16}{\volt}~\REF{AACMicrotec2016}.
A VPDHS OBDH system has been selected for the spacecraft operation, with a storage capacity of \SI{4}{\giga \byte}, and a power consumption of \SI{15}{\watt}~\REF{Digilent2016}.

\subsection{Propulsion}
\label{sec:Spacecraft_Propulsion}

The propulsion subsystem shall perform orbit injection, correction manoeuvres, detumbling, station keeping, collision avoidance, and de-orbiting (with $\Delta v$ values indicated in~\Table{tab:DeltaV Budget}).
A trade-off analysis to select a suitable type and quantity of thrusters was performed.
A Burn-Up/Break-Up (BUBU) manoeuvre for the disposal phase was taken into account, considering the mass of the spacecraft.

Driven by the mission's lifetime, a monopropellant hydrazine thruster design has been selected.
Four thrusters are used, in an arrangement that ensures thrust symmetry without changing the centre of gravity.
Each engine has a nominal thrust of \SI{20}{\newton} and a specific impulse of \SI{230}{\second}~\REF{ArianeGroup2016}.

The total mass of propellant was computed to be \SI{71.5}{\kilo \gram}, using Tsiolkovsky equation.
This amount of propellant needs a tank of about \SI{74}{\litre}, and thus the selected on is the Surface Tension tank OST 31/0~\REF{ArianeGroup2016_Tank}.

Drag perturbations shall be compensated by the electrical propulsion (FEEP thrusters), which are a part of the AOCS system.

\subsection{Mass \& power budget}
\label{sec:Mass_Power_Budgets}

With all subsystem information, plus the payload, the overall mass and power budget for the spacecraft has been built and is shown in~\Table{tab:MassPower_Budget}.
\begin{table}[!htbp]
    \footnotesize
    \centering
    \caption{Overall mass and power budget of the spacecraft, with margins for mass.}
    \label{tab:MassPower_Budget}
    \begin{tabular}{ccc}
    \toprule
        \textbf{Subsystem}  & \textbf{Mass [\si{kg}]}   & \textbf{Power [\si{W}]} \\
    \midrule
        Payload (+35\%)     & 289                       & 454           \\
        Telecom (+20\%)     & 23                        & 66            \\
        OBDH (+20\%)        & 3                         & 18            \\
        Power (+20\%)       & 29                        & 27            \\
        AOCS (+20\%)        & 84                        & 266           \\
        Propulsion (+20\%)  & 37                        & 24            \\
        Thermal (+20\%)     & 121                       & 36            \\
        Structure (+20\%)   & 350                       & 0             \\
        Propellant (+20\%)  & 71                        & 0             \\[0.8ex]
        \textbf{Total}      & \textbf{\num{1007}}       & \textbf{891}  \\
    \bottomrule
    \end{tabular}
\end{table}

Considering this mass and power budget, instead of developing a new satellite bus an existing one could be used, such as Astrosat-1000 from Airbus~\REF{Damilano2001,Gleyzes2012}.
This bus was used for the Pleiades constellation of remote sensing satellites, and can accommodate a total mass between \SIlist{800;1200}{\kilo \gram}.
The attitude control requirements of Magrathea are, in fact, less demanding than those of Pleiades.


%% file: 7_Programmatic.tex
%
%
%

\section{Programmatic}
\label{sec:Programmatic}

\subsection{Cost analysis}
\label{sec:Programmatic_Cost}

With the information gathered for the spacecraft subsystems and payload, a preliminary cost analysis was performed (see~\Table{tab:PCost}).
As a first approximation, it is estimated that Magrathea has a cost compatible with a M-class Cosmic Vision mission.
Since the project is still in an early-design phase, and several technologies still need to be developed (especially for some payload equipments), a 10\% margin is applied to the overall cost of the mission.

While spacecraft operations and mission management can be financed by a supporting agency, the payload costs should be charged to the national agencies and laboratories from the scientific consortium, since there is still much research to be performed.

\begin{table}[!htbp]
    \footnotesize
    \centering
    \caption{Preliminary cost analysis of the Magrathea mission.}
    \label{tab:PCost}
    \begin{tabular}{cc}
    \toprule
        \textbf{Spacecraft Elements}       & Cost [\si{M\EUR}] \\
    \midrule
        Payload                            & 88 \\
        Spacecraft bus                     & 135\\[0.4ex]
		\textit{Total}                     & \textit{223} \\
    \rule{0pt}{3ex} 
        \textbf{Mission and programmatic elements}   &   \\
    \midrule
        ESA program level                  & 27 \\
        Integration, assembly and test     & 22 \\
        Ground operations                  & 31 \\
        Flight software                    & 20 \\
        Launch vehicle                     & 75 \\[0.8ex]
        %
        \textbf{Total (10\% margin)}       & \textbf{438} \\
    \bottomrule
    \end{tabular}
\end{table}

\subsection{Risk assessment}
\label{sec:Programmatic_Risk}

Even though this is a preliminary design, an effort was made to identified possible risks, their consequences, and how they could be mitigated (shown in~\Table{tab:PRiskA}).
A classification of each risk, in terms of probability and impact, was also performed.

With this analysis it is possible to infer that the experimental volume and sample handling system are the most critical components (risks 1, 2, 9 and 10), as expected.
These are single points of failure, and should be extensively tested before launch (\textit{e.g.} using Failure mode, effects and criticality analysis, FMECA).

\begin{table*}[htbp]
    \scriptsize
    \renewcommand{\arraystretch}{1.2}
    \centering
    \caption{Preliminary risk assessment for the Magrathea mission, classified and with mitigation options.}
    \label{tab:PRiskA}
    \begin{tabular}{>{\centering}m{0.8cm} m{4cm} m{4cm} >{\centering}m{1cm} >{\centering}m{1cm} m{4.5cm}}
     \toprule
        \textbf{Risk ID}    & \textbf{Risk}
                            & \textbf{Consequence}
                            & \textbf{Probability} & \textbf{Impact}
                            & \textbf{Mitigation} \\
     \midrule
        \underline{R1}      & Experimental volume or sampling handle systems fail.
                            & No further experiments possible (reduced science return).
                            & \textbf{High}        & \textit{Medium}
                            & FMECA of chamber, to reduce probability of failure, and verification by tests.\\
        \underline{R2}      & Experimental volume contamination cannot be effectively controlled.
                            & Chamber lifetime limited (reduced science return).
                            & \textbf{High}         & \textit{Medium}
                            & Prioritise experiments by lowest contamination risk, and reduce their number as possible. Verify chamber operation.\\
        R3                  & Louvre blocking.
                            & Thermal instability on payload section.
                            & \textit{Medium}       & \textit{Medium}
                            & Compensate with active control of other louvres.\\
        R4                  & Primary thermal control failure.
                            & Thermal instability on bus section.
                            & \textit{Medium}       & Low
                            & Compensation with heaters and louvre operation.\\
        R5                  & PDCU malfunction.
                            & Partial or total loss of power.
                            & Low                   & \textit{Medium}
                            & Triple redundancy on electronics boards.\\
        R6                  & Debris impact on solar array/sunshield.
                            & Lower power production and thermal instability.
                            & \textit{Medium}       & Low
                            & Collision avoidance manoeuvres, and structure reinforcement.\\
        R7                  & Unplanned and/or uncontrolled venting of experimental volume.
                            & Lost of pointing and orbit.
                            & Low                   & \textit{Medium}
                            & Controlled venting procedure in place with extensive test.\\
        R8                  & Water condensation during heating.
                            & Unrepresentative conditions created in experimental volume.
                            & Low                   & \textit{Medium}
                            & Maintain low pressure in experimental volume.\\
        \underline{\textbf{R9}}         & Particles exceed maximum allowable velocity after agitation manoeuvre.
                            & Unrepresentative conditions created, no observation of the collisions, lower growth probability for small particles (reduced science return).
                            & \textbf{High}         & \textbf{High}
                            & Extensive test of agitation control mechanisms.\\
        \underline{R10}     & Too many particles stick to the walls of the experimental volume.
                            & No observations of collisions possible (reduced science return).
                            & \textbf{High}         & \textit{Medium}
                            & Extensive test of agitation control mechanisms and material properties.\\
     \bottomrule
    \end{tabular}
\end{table*}



\subsection{Outreach}
\label{sec:Outreach}

Every mission nowadays should prepare an outreach campaign.
This not only publicises the mission itself, but promotes all space activities and their significance on people daily lives.
Furthermore, by publicising space science and technology to a new generation, the mission can contribute to an increase in future space jobs.
As this project was developed within the Alpbach Summer School, an activity sponsored by ESA, the outreach activities have been tailored to ESA's education office ones~\REF{ESAAcademy2017}.

Magrathea presents several opportunities for outreach and hands-on projects.
For example, images of particle agglomeration could be used to teach children and students how planets are thought to form.
Another option, for more advanced students, is to expand upon ESA's thesis opportunities, such as Drop Your Thesis and Fly Your Thesis.
A new category, Orbit Your Thesis, could be introduced, in which, one or more sample tubes could be assigned as the prize for a competition.
This would allow for university students from ESA member states to choose the particles and experiment conditions to be used for some of the experiments.


%% file: 8_Conclusion.tex
%
%
%

\section{Final remarks}
\label{sec:Conclusions}

Magrathea will help to answer one of the biggest questions in astrophysics and a decadal survey priority~\REF{NationalResearchCouncil2010}: how planets form from dust in protoplanetary disks.
The mission will advance our understanding of the physics underlying dust grain growth from the microscopic to the macroscopic scale.
Twenty eight experiments will study the influence of a range of variables, including grain size, composition, and collision speed, on grain growth.
A set of science instruments, including cameras, an optical microscope, and an atomic force microscope will monitor particle growth.

Beyond astrophysics, the physics studied by Magrathea are relevant to many other granular processes such as industrial powder processes.
Studying them under a microgravity environment removes the complicating effects of Earth's gravity, enabling a clearer view of the grain growth process, and ultimately improving the efficiency of manufacturing.
In addition, the mission includes opportunities for student participation and a wide range of public outreach opportunities.

Magrathea is a first of its kind orbiting laboratory in low Earth orbit.
No other microgravity grain growth experiment has been developed with such a large experimental volume and long duration.
This configuration is the next logical step in studying dust grain growth, and builds upon a strong heritage of drop tower experiments, parabolic flight, and ISS-based experiments.
The spacecraft bus can be derived from one of the existing industry buses (\textit{e.g.} Astrosat-1000).
As a result the mission provides high science return for relatively low risk and cost.




%% file: GreenTeamPaper_ArXiv.bbl
\begin{thebibliography}{10}
\expandafter\ifx\csname url\endcsname\relax
  \def\url#1{\texttt{#1}}\fi
\expandafter\ifx\csname urlprefix\endcsname\relax\def\urlprefix{URL }\fi
\expandafter\ifx\csname href\endcsname\relax
  \def\href#1#2{#2} \def\path#1{#1}\fi

\bibitem{Borucki2010}
W.~J. Borucki, D.~Koch, G.~Basri, N.~Batalha, T.~Brown, D.~Caldwell,
  J.~Caldwell, J.~Christensen-Dalsgaard, W.~D. Cochran, E.~DeVore, E.~W.
  Dunham, A.~K. Dupree, T.~N. Gautier, J.~C. Geary, R.~Gilliland, A.~Gould,
  S.~B. Howell, J.~M. Jenkins, Y.~Kondo, D.~W. Latham, G.~W. Marcy, S.~Meibom,
  H.~Kjeldsen, J.~J. Lissauer, D.~G. Monet, D.~Morrison, D.~Sasselov,
  J.~Tarter, A.~Boss, D.~Brownlee, T.~Owen, D.~Buzasi, D.~Charbonneau,
  L.~Doyle, J.~Fortney, E.~B. Ford, M.~J. Holman, S.~Seager, J.~H. Steffen,
  W.~F. Welsh, J.~Rowe, H.~Anderson, L.~Buchhave, D.~Ciardi, L.~Walkowicz,
  W.~Sherry, E.~Horch, H.~Isaacson, M.~E. Everett, D.~Fischer, G.~Torres, J.~A.
  Johnson, M.~Endl, P.~MacQueen, S.~T. Bryson, J.~Dotson, M.~Haas,
  J.~Kolodziejczak, J.~{Van Cleve}, H.~Chandrasekaran, J.~D. Twicken, E.~V.
  Quintana, B.~D. Clarke, C.~Allen, J.~Li, H.~Wu, P.~Tenenbaum, E.~Verner,
  F.~Bruhweiler, J.~Barnes, A.~Prsa, {Kepler Planet-Detection Mission:
  Introduction and First Results}, Science 327~(5968) (2010) 977--980.
\newblock \href {http://dx.doi.org/10.1126/science.1185402}
  {\path{doi:10.1126/science.1185402}}.

\bibitem{Broeg2013}
C.~Broeg, A.~Fortier, D.~Ehrenreich, Y.~Alibert, W.~Baumjohann, W.~Benz,
  M.~Deleuil, M.~Gillon, A.~Ivanov, R.~Liseau, M.~Meyer, G.~Oloffson,
  I.~Pagano, G.~Piotto, D.~Pollacco, D.~Queloz, R.~Ragazzoni, E.~Renotte,
  M.~Steller, N.~Thomas, {CHEOPS: A transit photometry mission for ESA's small
  mission programme}, EPJ Web of Conferences 47 (2013) 03005.
\newblock \href {http://dx.doi.org/10.1051/epjconf/20134703005}
  {\path{doi:10.1051/epjconf/20134703005}}.

\bibitem{Knutson2007}
H.~A. Knutson, {Extrasolar planets: Water on distant worlds}, Nature 448~(7150)
  (2007) 143--145.
\newblock \href {http://dx.doi.org/10.1038/448143a}
  {\path{doi:10.1038/448143a}}.

\bibitem{Testi2014}
L.~Testi, T.~Birnstiel, L.~Ricci, S.~Andrews, J.~Blum, J.~Carpenter,
  C.~Dominik, A.~Isella, A.~Natta, J.~P. Williams, D.~J. Wilner, {Dust
  Evolution in Protoplanetary Disks}, in: Protostars and Planets VI, University
  of Arizona Press, 2014.
\newblock \href {http://dx.doi.org/10.2458/azu_uapress_9780816531240-ch015}
  {\path{doi:10.2458/azu_uapress_9780816531240-ch015}}.

\bibitem{Blum1996}
J.~Blum, G.~Wurm, S.~Kempf, T.~Henning, {The Brownian Motion of Dust Particles
  in the Solar Nebula: An Experimental Approach to the Problem of Pre-planetary
  Dust Aggregation}, Icarus 124~(2) (1996) 441--451.
\newblock \href {http://dx.doi.org/10.1006/icar.1996.0221}
  {\path{doi:10.1006/icar.1996.0221}}.

\bibitem{Zsom2010}
A.~Zsom, C.~W. Ormel, C.~G{\"{u}}ttler, J.~Blum, C.~P. Dullemond, {The outcome
  of protoplanetary dust growth: pebbles, boulders, or planetesimals? II.
  Introducing the bouncing barrier}, Astronomy and Astrophysics 513 (2010) A57.
\newblock \href {http://dx.doi.org/10.1051/0004-6361/200912976}
  {\path{doi:10.1051/0004-6361/200912976}}.

\bibitem{Blum2008}
J.~Blum, G.~Wurm, {The Growth Mechanisms of Macroscopic Bodies in
  Protoplanetary Disks}, Annual Review of Astronomy and Astrophysics 46~(1)
  (2008) 21--56.
\newblock \href {http://dx.doi.org/10.1146/annurev.astro.46.060407.145152}
  {\path{doi:10.1146/annurev.astro.46.060407.145152}}.

\bibitem{Guttler2010}
C.~G{\"{u}}ttler, J.~Blum, A.~Zsom, C.~W. Ormel, C.~P. Dullemond, {The outcome
  of protoplanetary dust growth: pebbles, boulders, or planetesimals? I.
  Mapping the zoo of laboratory collision experiments}, Astronomy and
  Astrophysics 513 (2010) A56.
\newblock \href {http://dx.doi.org/10.1051/0004-6361/200912852}
  {\path{doi:10.1051/0004-6361/200912852}}.

\bibitem{Brisset2015}
J.~Brisset, J.~Colwell, A.~Dove, D.~Maukonen, N.~Brown, K.~Lai, B.~Hoover,
  {NanoRocks: Studying Planet Formation and Planetary Rings on the
  International Space Station}, European Planetary Science Congress 10 (2015)
  EPSC2015--767.

\bibitem{NASA_droptower}
{National Aeronautics and Space Administration},
  \href{www1.grc.nasa.gov/facilities/zero-g/}{{The Zero Gravity Research
  Facility}}, visited on 23rd of July 2017 (2017).
\newline\urlprefix\url{www1.grc.nasa.gov/facilities/zero-g/}

\bibitem{Blum2014}
J.~Blum, E.~Beitz, M.~Bukhari, B.~Gundlach, J.-H. Hagemann, D.~Hei{\ss}elmann,
  S.~Kothe, R.~Schr{\"{a}}pler, I.~von Borstel, R.~Weidling, {Laboratory Drop
  Towers for the Experimental Simulation of Dust-aggregate Collisions in the
  Early Solar System}, Journal of Visualized Experiments~(88).
\newblock \href {http://dx.doi.org/10.3791/51541} {\path{doi:10.3791/51541}}.

\bibitem{Blum2001}
J.~Blum, G.~Wurm, {Drop tower experiments on sticking, restructuring, and
  fragmentation of preplanetary dust aggregates}, Microgravity Science and
  Technology 13~(1) (2001) 29--34.
\newblock \href {http://dx.doi.org/10.1007/BF02873329}
  {\path{doi:10.1007/BF02873329}}.

\bibitem{Brisset2013}
J.~Brisset, D.~Hei{\ss}elmann, S.~Kothe, R.~Weidling, J.~Blum, {The suborbital
  particle aggregation and collision experiment (SPACE): Studying the collision
  behavior of submillimeter-sized dust aggregates on the suborbital rocket
  flight REXUS 12}, Review of Scientific Instruments 84~(9) (2013) 094501.
\newblock \href {http://dx.doi.org/10.1063/1.4819443}
  {\path{doi:10.1063/1.4819443}}.

\bibitem{NASA_sounding}
{National Aeronautics and Space Administration},
  \href{eea.spaceflight.esa.int/portal/exp/?id=7955}{{Cosmic dust aggregation
  experiment}}, visited on 23rd of July 2017 (2017).
\newline\urlprefix\url{eea.spaceflight.esa.int/portal/exp/?id=7955}

\bibitem{Love2014}
S.~G. Love, D.~R. Pettit, S.~R. Messenger, {Particle aggregation in
  microgravity: Informal experiments on the International Space Station},
  Meteoritics {\&} Planetary Science 49~(5) (2014) 732--739.
\newblock \href {http://dx.doi.org/10.1111/maps.12286}
  {\path{doi:10.1111/maps.12286}}.

\bibitem{Colwell2017}
J.~Colwell, J.~Brisset, A.~Dove, L.~Roe, J.~Blum, {Q-PACE: the CubeSat Particle
  Aggregation and Collision Experiment}, in: 12th Low Cost Planetary Missions
  Conference, 2017.

\bibitem{Pollack1985}
J.~B. Pollack, C.~P. McKay, B.~M. Christofferson, {A calculation of the
  Rosseland mean opacity of dust grains in primordial solar system nebulae},
  Icarus 64~(3) (1985) 471--492.
\newblock \href {http://dx.doi.org/10.1016/0019-1035(85)90069-7}
  {\path{doi:10.1016/0019-1035(85)90069-7}}.

\bibitem{Keller2002}
L.~P. Keller, S.~Hony, J.~P. Bradley, F.~J. Molster, L.~B. F.~M. Waters,
  J.~Bouwman, A.~de~Koter, D.~E. Brownlee, G.~J. Flynn, T.~Henning,
  H.~Mutschke, {Identification of iron sulphide grains in protoplanetary
  disks}, Nature 417~(6884) (2002) 148--150.
\newblock \href {http://dx.doi.org/10.1038/417148a}
  {\path{doi:10.1038/417148a}}.

\bibitem{Dominik1997}
C.~Dominik, A.~G. G.~M. Tielens, {The Physics of Dust Coagulation and the
  Structure of Dust Aggregates in Space}, The Astrophysical Journal 480~(2)
  (1997) 647--673.
\newblock \href {http://dx.doi.org/10.1086/303996} {\path{doi:10.1086/303996}}.

\bibitem{Wada2007}
K.~Wada, H.~Tanaka, T.~Suyama, H.~Kimura, T.~Yamamoto, {Numerical Simulation of
  Dust Aggregate Collisions. I. Compression and Disruption of Two‐Dimensional
  Aggregates}, The Astrophysical Journal 661~(1) (2007) 320--333.
\newblock \href {http://dx.doi.org/10.1086/514332} {\path{doi:10.1086/514332}}.

\bibitem{Wada2008}
K.~Wada, H.~Tanaka, T.~Suyama, H.~Kimura, T.~Yamamoto, {Numerical Simulation of
  Dust Aggregate Collisions. II. Compression and Disruption of
  Three‐Dimensional Aggregates in Head‐on Collisions}, The Astrophysical
  Journal 677~(2) (2008) 1296--1308.
\newblock \href {http://dx.doi.org/10.1086/529511} {\path{doi:10.1086/529511}}.

\bibitem{Rumble2017}
J.~R. Rumble (Ed.), {CRC Handbook of Chemistry and Physics}, 98th Edition, CRC
  Press, 2017.

\bibitem{Heiselmann2007}
D.~Hei{\ss}elmann, H.~J. Fraser, J.~Blum, {Experimental Studies on the
  Aggregation Properties of Ice and Dust in Planet-Forming Regions}, in:
  Procedings of the 58th International Astronautical Congress, Paris, 2007, pp.
  IAC--07--A2.1.02.
\newblock \href {http://arxiv.org/abs/1106.4760} {\path{arXiv:1106.4760}}.

\bibitem{Arianespace2012}
Arianespace,
  \href{http://www.arianespace.com/wp-content/uploads/2015/09/Soyuz-Users-Manual-March-2012.pdf}{{Soyuz
  User's Manual, Issue 2 - Revision 0}}, Tech. rep., retrieved on 23rd of July
  2017 (2012).
\newline\urlprefix\url{http://www.arianespace.com/wp-content/uploads/2015/09/Soyuz-Users-Manual-March-2012.pdf}

\bibitem{Bibring2007}
J.-P. Bibring, P.~Lamy, Y.~Langevin, A.~Soufflot, M.~Berth{\'{e}}, J.~Borg,
  F.~Poulet, S.~Mottola, {CIVA}, Space Science Reviews 128~(1-4) (2007)
  397--412.
\newblock \href {http://dx.doi.org/10.1007/s11214-006-9135-5}
  {\path{doi:10.1007/s11214-006-9135-5}}.

\bibitem{riedler2007midas}
W.~Riedler, K.~Torkar, H.~Jeszenszky, J.~Romstedt, H.~S.~C. Alleyne, H.~Arends,
  W.~Barth, J.~V.~D. Biezen, B.~Butler, P.~Ehrenfreund, M.~Fehringer,
  G.~Fremuth, J.~Gavira, O.~Havnes, E.~K. Jessberger, R.~Kassing,
  W.~Kl{\"{o}}ck, C.~Koeberl, A.~C. Levasseur-Regourd, M.~Maurette,
  F.~R{\"{u}}denauer, R.~Schmidt, G.~Stangl, M.~Steller, I.~Weber, {MIDAS –
  The Micro-Imaging Dust Analysis System for the Rosetta Mission}, Space
  Science Reviews 128~(1-4) (2007) 869--904.
\newblock \href {http://dx.doi.org/10.1007/s11214-006-9040-y}
  {\path{doi:10.1007/s11214-006-9040-y}}.

\bibitem{chobotov1993}
M.~V. Chobotov, G.~P. Purohit, {Low-gravity propellant gauging system for
  accurate predictions of spacecraft end-of-life}, Journal of Spacecraft and
  Rockets 30~(1) (1993) 92--101.
\newblock \href {http://dx.doi.org/10.2514/3.25475}
  {\path{doi:10.2514/3.25475}}.

\bibitem{neubauer1993}
F.~Neubauer, H.~Marschall, M.~Pohl, K.-H. Glassmeier, G.~Musmann, F.~Mariani,
  M.~Acuna, L.~Burlaga, N.~Ness, M.~Wallis, et~al., {First results from the
  Giotto magnetometer experiment during the P/Grigg-Skjellerup encounter},
  Astronomy and Astrophysics 268 (1993) L5--L8.

\bibitem{Iafolla2010}
V.~Iafolla, E.~Fiorenza, C.~Lefevre, A.~Morbidini, S.~Nozzoli, R.~Peron,
  M.~Persichini, A.~Reale, F.~Santoli, {Italian Spring Accelerometer (ISA): A
  fundamental support to BepiColombo Radio Science Experiments}, Planetary and
  Space Science 58~(1-2) (2010) 300--308.
\newblock \href {http://dx.doi.org/10.1016/j.pss.2009.04.005}
  {\path{doi:10.1016/j.pss.2009.04.005}}.

\bibitem{AMRPropulsionTechnologies2017}
{AMR Propulsion Technologies}, \href{www.enpulsion.com}{{Data Sheet -
  Introduction to the IFM Nano Thruster}}, retrieved on 25th of July 2017
  (2017).
\newline\urlprefix\url{www.enpulsion.com}

\bibitem{NorthropGrumman2017}
{Northrop Grumman},
  \href{www.northropgrumman.com/Capabilities/LN200sInertial/Pages/default.aspx}{{Data
  Sheet - LN-200S Inertial Measurement Unit}}, retrieved on 25th of July 2017
  (2017).
\newline\urlprefix\url{www.northropgrumman.com/Capabilities/LN200sInertial/Pages/default.aspx}

\bibitem{SinclairInterplanetary2016}
{Sinclair Interplanetary},
  \href{www.sinclairinterplanetary.com/startrackers}{{Data Sheet - Second
  Generation Star Tracker (ST-16RT2)}}, retrieved on 25th of July 2017 (2016).
\newline\urlprefix\url{www.sinclairinterplanetary.com/startrackers}

\bibitem{NewSpaceSystems2016}
{NewSpace Systems}, \href{www.newspacesystems.com}{{Data Sheet - Fine (Digital)
  Sun Sensor}}, retrieved on 25th of July 2017 (2016).
\newline\urlprefix\url{www.newspacesystems.com}

\bibitem{AirbusDefense&Space2016}
{Airbus Defense {\&} Space},
  \href{spaceequipment.airbusdefenceandspace.com/avionics/control-momentum-gyroscopes/cmg-15-45s/}{{Data
  Sheet - CMG 15-45S}}, retrieved on 25th of July 2017 (2016).
\newline\urlprefix\url{spaceequipment.airbusdefenceandspace.com/avionics/control-momentum-gyroscopes/cmg-15-45s/}

\bibitem{NewSpaceSystems2016a}
{NewSpace Systems}, \href{www.newspacesystems.com}{{Data Sheet - Magnetorquer
  Rod}}, retrieved on 25th of July 2017 (2016).
\newline\urlprefix\url{www.newspacesystems.com}

\bibitem{SurreySatelliteTechnology2016}
{Surrey Satellite Technology},
  \href{www.sst-us.com/shop/satellite-subsystems/global-positioning-systems-gps-receivers/sgr-20-space-gps-receiver}{{Data
  Sheet - SGR-20}}, retrieved on 25th of July 2017 (2016).
\newline\urlprefix\url{www.sst-us.com/shop/satellite-subsystems/global-positioning-systems-gps-receivers/sgr-20-space-gps-receiver}

\bibitem{SpectrolabInc.2010}
{Spectrolab Inc.}, \href{http://www.spectrolab.com/solarpanels.htm}{{Data Sheet
  - Space Solar Panels}}, retrieved on 23rd of July 2017 (2010).
\newline\urlprefix\url{http://www.spectrolab.com/solarpanels.htm}

\bibitem{ASP-EquipmentGmbH2017}
{ASP-Equipment GmbH},
  \href{https://www.asp-equipment.de/products/space-products/detail/news/power-control-unit-with-maximum-power-point-tracker/?tx{\_}news{\_}pi1{\%}5Bcontroller{\%}5D=News{\&}tx{\_}news{\_}pi1{\%}5Baction{\%}5D=detail{\&}cHash=0c49daba943e3f82d6c7f971f5a39f2c}{{Data
  Sheet - Power Control Unit with Maximum Power Point Tracker}}, retrieved on
  23rd of July 2017 (2017).
\newline\urlprefix\url{https://www.asp-equipment.de/products/space-products/detail/news/power-control-unit-with-maximum-power-point-tracker/?tx{\_}news{\_}pi1{\%}5Bcontroller{\%}5D=News{\&}tx{\_}news{\_}pi1{\%}5Baction{\%}5D=detail{\&}cHash=0c49daba943e3f82d6c7f971f5a39f2c}

\bibitem{EaglePicherTechnologiesLLC2011}
{EaglePicher Technologies LLC},
  \href{https://www.eaglepicher.com/technology/battery-chemistries/lithium-ion/}{{Data
  Sheet - EaglePicher SLC-028-01 Cell}}, retrieved on 23rd of July 2017 (2011).
\newline\urlprefix\url{https://www.eaglepicher.com/technology/battery-chemistries/lithium-ion/}

\bibitem{EnduroSat2016}
EnduroSat,
  \href{www.endurosat.com/products/cubesat-s-band-patch-antenna/}{{User Manual
  - S-Band Patch Antenna Type I}}, retrieved on 23rd of July 2017 (2016).
\newline\urlprefix\url{www.endurosat.com/products/cubesat-s-band-patch-antenna/}

\bibitem{Elta2016}
Elta,
  \href{www.elta.fr/gb/products_balloon/product/s-band-transceivers-en.html}{{Data
  Sheet - S-Band Transceiver}}, retrieved on 23rd of July 2017 (2016).
\newline\urlprefix\url{www.elta.fr/gb/products_balloon/product/s-band-transceivers-en.html}

\bibitem{A-Info2016}
A-Info, \href{www.ainfoinc.com/en/p_ant_h_std.asp}{{Data Sheet - LB-90-10
  Standard Gain Horn Antenna}}, retrieved on 23rd of July 2017 (2016).
\newline\urlprefix\url{www.ainfoinc.com/en/p_ant_h_std.asp}

\bibitem{AACMicrotec2016}
A.~Microtec,
  \href{aacmicrotec.com/products/spacecraft-subsystems/avionics/siriuscdh/}{{Data
  Sheet - Sirius C{\&}DH}}, retrieved on 19th of May 2017 (2016).
\newline\urlprefix\url{aacmicrotec.com/products/spacecraft-subsystems/avionics/siriuscdh/}

\bibitem{Digilent2016}
V.~Aerospace,
  \href{www.vectronic-aerospace.com/space-applications/payload-data-handling-system-vpdhs/}{{Data
  Sheet - OBDH VPDHS}}, retrieved on 19th of May 2017 (2016).
\newline\urlprefix\url{www.vectronic-aerospace.com/space-applications/payload-data-handling-system-vpdhs/}

\bibitem{ArianeGroup2016}
A.~S. Lauchers,
  \href{www.space-propulsion.com/spacecraft-propulsion/hydrazine-thrusters/20n-hydrazine-thruster.html}{{Data
  Sheet - 20N Monopropellant Hydrazine Thruster}}, retrieved on 19th of May
  2017 (2016).
\newline\urlprefix\url{www.space-propulsion.com/spacecraft-propulsion/hydrazine-thrusters/20n-hydrazine-thruster.html}

\bibitem{ArianeGroup2016_Tank}
A.~S. Lauchers,
  \href{www.space-propulsion.com/spacecraft-propulsion/hydrazine-tanks/index.html}{{Data
  Sheet - Model OST 31-0}}, retrieved on 19th of May 2017 (2016).
\newline\urlprefix\url{www.space-propulsion.com/spacecraft-propulsion/hydrazine-tanks/index.html}

\bibitem{Damilano2001}
P.~Damilano, {Pleiades High Resolution Satellite: A Solution for Military and
  Civilian Needs in Metric-Class Optical Observation}, in: 15th Annual AIAA/USU
  Conference on Small Satellites, 2001.

\bibitem{Gleyzes2012}
M.~A. Gleyzes, L.~Perret, P.~Kubik, {PLEIADES SYSTEM ARCHITECTURE AND MAIN
  PERFORMANCES}, ISPRS - International Archives of the Photogrammetry, Remote
  Sensing and Spatial Information Sciences XXXIX-B1 (2012) 537--542.
\newblock \href {http://dx.doi.org/10.5194/isprsarchives-XXXIX-B1-537-2012}
  {\path{doi:10.5194/isprsarchives-XXXIX-B1-537-2012}}.

\bibitem{ESAAcademy2017}
{European Space Agency}, \href{www.esa.int/Education/ESA{\_}Academy}{{ESA
  Academy / Education / ESA}}, visited on 12th of November 2017 (2017).
\newline\urlprefix\url{www.esa.int/Education/ESA{\_}Academy}

\bibitem{NationalResearchCouncil2010}
{National Research Council}, {New Worlds, New Horizons in Astronomy and
  Astrophysics}, National Academies Press, Washington, D.C., 2010.
\newblock \href {http://dx.doi.org/10.17226/12951} {\path{doi:10.17226/12951}}.

\end{thebibliography}
